\documentstyle[11pt]{article}
\begin{document}

\title{ Top quark pair production via polarized and unpolarized photons in
        Supersymmetric QCD}

\author{{
  Han Liang$^{b,}$\footnote{email:rdhan@hpe25.nic.ustc.edu.cn},
  Ma Wen-Gan$^{a,b}$, Yu Zeng-hui$^{b}$}\\
{\small $^{a}$CCAST (World Laboratory), P.O.Box 8730, Beijing 100080,P.R.China}\\
{\small $^{b}$Department of Modern Physics, University of Science and Technology}\\
{\small of China (USTC), Hefei, Anhui 230027, P.R.China, \footnote{Mailing address}}
}
\date{}
\maketitle

\vskip 12mm

\begin{center}\begin{minipage}{5in}

\begin{center} ABSTRACT\end{center}
\baselineskip 0.3in

{QCD corrections to top quark pair production via fusion of both
polarized and unpolarized photons are calculated in Supersymmetric
Model. The corrections are found to be sizable. The dependence of
the corrections on the masses of the supersymmetric particles is
also investigated. Furthermore, we studied CP asymmetry effects
arising from the complex couplings in the MSSM. The CP violating
parameter can reach $10^{-2}$ for favorable parameter values.} \\

\vskip 10mm

{~~~~PACS number(s): 13.65.+i, 13.88.+e, 14.65.-q, 14.80.Dq, 14.80.Gt}
\end{minipage}
\end{center}

\baselineskip=0.36in

\newpage
\noindent
\begin{large}
\baselineskip 0.35in
\begin{flushleft} {\bf I. Introduction} \end{flushleft}

   Direct evidence for the top quark was recently presented by the CDF
(Collider Detector at Fermilab) Collaboration \cite{a1}. This is considered
to be a remarkable success for the Standard Model (SM), since the value of
the top mass determined by them, $174\pm 10^{+13}_{-12}~GeV$, coincides with
the indirect determination from the available precise data of electroweak
experiments. Due to the determination of the top quark mass, it is certain
that a number of new and interesting issues in top quark pair production
and decay will arise. At the Next Linear Collider (NLC), operating in
photon-photon collision mode at a design energy of $500 \sim 2000~GeV$
with a luminosity of the order $10^{33} cm^{-2} s^{-1}$ \cite{a2}, a large
number of heavy quark pairs and other new particles can be produced with
an agreeable production rate \cite{a3} . Those events would be much cleaner
than those produced at $pp$ and $p \bar{p}$ colliders. It has been also
found that the $t\bar{t}$ production rate in $\gamma\gamma$ collisions
from laser back-scattering is much larger than that from the direct
$e^{+}e^{-} \rightarrow t\bar{t}$ production both with and without considering
the threshold QCD effect, for $m_t \sim 100~-~200 GeV$ at center-of-mass
energies of the electron-positron system around 1 TeV.\cite{a4}. Thus the
process $\gamma\gamma \rightarrow t \bar{t}$ has a large potential for
producing and studying heavy quarks directly. The next-to-leading order
QCD corrections in the SM for this process both for polarized and unpolarized
particle collisions have been discussed in detail in Ref.\cite{a5}. There
it was shown that these QCD corrections are significant, about $10 \%$.
However, the process also seems to be sensitive in searching for virtual
effects from new particles, which would imply physics beyond the Standard
Model. Among various models which introduce new physics, the Minimal
Supersymmetric Standard Model (MSSM)\cite{a6} is the most attractive one
at present, since it is the simplest case of Supersymmetric Models (SUSY).
Therefore it is important to study the physical effects induced by the
virtual supersymmetric particles for top quark pair production via photon
fusion. If we assume that one of the stop quarks are much lighter than
the other supersymmetric quarks due to strong mixing, the process
$\gamma\gamma \rightarrow t\bar{t}$ would be a very interesting reaction for
testing supersymmetric theory in that the corrections due to virtual
$\tilde{t}$ would provide evidence for the existence of the quark scalar
partners. Recently H. Wang el al. have calculated the Supersymmetric QCD
corrections for this process via unpolarized photon collisions in some
cases in Ref.\cite{a7}. \\

  With the advent of new collider techniques, back-scattered laser photons
can be linearly polarized with high luminosity and rather high efficiency
\cite{a8}. These techniques will undoubtly be used to improve our knowledge
of the top quark parameters, particularly when one combines the data
from top pair production via photon-photon fusion with other data from
$e^{+}e^{-}$ and $pp$ collisions.
In addition to that general advantage, the process of top quark pair produced
by polarized photon collisions may provide us with a better understanding of
CP-violation phenomena \cite{a9}. Since the mixing of top quarks with
other generations is very small, CP-violation in this process is negligibly
small within the Kobayashi-Maskawa framework of the Standard Model.
However, in the MSSM, even without generation mixing, there are more
possibilities to introduce complex couplings than in the SM.
Here in this paper, we limit ourselves only to the combination of the
CP-violating phases $\phi_{\tilde{t}}$ and $\phi_{\tilde{g}}$, which emerge in
the stop quark mixing matrix and in the Majorana mass term of the gluino
\cite{a10}. These non-zero complex phases, which occur in the Lagrangian, cannot
be rotated away by a suitable redefinition of the fields. They will lead to CP
violation within a single generation. Therefore, the observation of CP-violation
in the process of top quark pair production via polarized $\gamma\gamma$ fusion
may help to promote our understanding of the features of the combination of
CP-violating phases $\phi_{\tilde{t}}$ and $\phi_{\tilde{g}}$.

The paper is organized as follows: in Sec. II, the renormalization of the top
quark self-energy in the frame of the MSSM is described, and the explicit
analytical forms of the Lorentz invariant matrix elements including the
next-to-leading order of supersymmetric QCD corrections are presented.
In Sec. III, the numerical results and discussions are depicted. Finally,
the conclusions are given. In the appendix, the form factors used for the
cross section calculations are listed.\\

\begin{flushleft} {\bf II. Calculations} \end{flushleft}
In this work, we denote the reaction discussed as

$$
\gamma (p_3, \lambda_1) \gamma (p_4, \lambda_2) \longrightarrow t(p_2) \bar{t}(p_1)
\eqno{(2.1)}
$$

where $\lambda_{1,2}=\pm 1$, $p_2$ and $p_1$ represent the momenta of the
outgoing top quark and its anti-particle, $p_3$ and $p_4$ are for the momenta
of the two incoming photons with helicities $\lambda_1$ and $\lambda_2$
respectively.

The Feynman diagrams for the process $\gamma\gamma \rightarrow t\bar{t}$ are
shown in Fig.1, and the relevant Feynman rules can be found in \cite{a6}. In
the calculation dimensional regularization and the on-mass-shell (OMS)
renormalization scheme\cite{a11} are adopted to eliminate all the ultraviolet
divergences. The interaction Lagrangian for gluino-stop-top coupling including
the CP-violating phases is given by

$$
\begin{array} {lll}
 L_{\tilde{g} \tilde{t} \bar{t}} & = & - \sqrt{2} g_s T^a ~\bar{t} ~
   [(cos \theta_{\tilde t} ~ \tilde{t}_1 +
     sin \theta_{\tilde t} ~ \tilde{t}_2) e^{-i \phi} P_{R} \\
 & & + (sin \theta_{\tilde t} ~ \tilde{t}_1 -
        cos \theta_{\tilde t} ~ \tilde{t}_2) e^{i \phi} P_{L}]
        ~ \tilde{g}_a + h.c. \\
 & = & - \sqrt{2} g_s T^a ~ \bar{t} ~
      [(A_{R} ~ \tilde{t}_1 + B_{R} ~ \tilde{t}_2) P_{R} +
       (A_{L} ~ \tilde{t}_1 - B_{L} ~ \tilde{t}_2) P_{L}]
       ~ \tilde{g}_a + h.c. \\
\end{array}
\eqno{(2.2)}
$$

where $g_s$ is the strong coupling constant, $T^a$ are $SU(3)_C$ generators
and $P_{R,L} = \frac{1}{2}(1 \pm \gamma_5)$. $\theta_{\tilde t}$ is the stop
mixing angle which transforms the stop mass eigenstates $\tilde{t}_n$ (n=1,2)
to the weak eigenstates $\tilde{t}_L$ and $\tilde{t}_R$.
$\phi = \phi_{\tilde{t}} - \phi_{\tilde{g}}$ is the combination of the two
phase angles. In the Equation (2.2) we denote

$$
\begin{array} {cc}
A_R = cos\theta_{\tilde t} e^{-i \phi}, & A_L = sin\theta_{\tilde t} e^{i \phi}, \\
B_R = sin\theta_{\tilde t} e^{-i \phi}, & B_L = cos\theta_{\tilde t} e^{i \phi} \\
\end{array}
\eqno{(2.3)}
$$

  With the introduction of the CP-violating phase $\phi$ in the MSSM, the
renormalized one-particle irreducible two-point function for top quarks
containing the contributions from SUSY QCD should be written as \cite{a12}

$$
\begin{array} {lll}
\hat{\Gamma}(p) &=& i (\rlap/p-m_{t}) + i \left [ \rlap/p P_{L}
                         \hat{\Sigma}^{L}(p^2)
                        + \rlap/p P_{R} \hat{\Sigma}^{R}(p^2)
                        + P_{L} \hat{\Sigma}^{S}(p^2)
                        + P_{R} \hat{\Sigma}^{S~\ddag}(p^2) \right].
\end{array}
\eqno{(2.4)}
$$

  It should be mentioned here that in the above equation the upper
conjugation symbol $\ddag$ acts only on the CP-violating phase
$\phi$. The corresponding unrenormalized top quark self energies read

$$
\begin{array} {lll}
\Sigma^{L}(p^2) &=& -\frac{C_F}{8 \pi^2} g_{s}^2
        \left ( A_{R} B_{L} B_{1}[p^2, m_{\tilde{g}}, m_{\tilde{t}_1}]
         + A_{L} B_{R} B_{1}[p^2, m_{\tilde{g}}, m_{\tilde{t}_2}] \right ),
\end{array}
\eqno{(2.5)}
$$

$$
\begin{array} {lll}
\Sigma^{R}(p^2) &=& -\frac{C_F}{8 \pi^2} g_{s}^2
        \left ( A_{L} B_{R} B_{1}[p^2, m_{\tilde{g}}, m_{\tilde{t}_1}]
         + A_{R} B_{L} B_{1}[p^2, m_{\tilde{g}}, m_{\tilde{t}_2}] \right ),
\end{array}
\eqno{(2.6)}
$$

$$
\begin{array} {lll}
\Sigma^{S}(p^2) &=& \frac{C_F}{8 \pi^2} g_{s}^2 m_{\tilde{g}} A_{L} B_{L}
        \left ( B_{0}[p^2, m_{\tilde{g}}, m_{\tilde{t}_1}]
         - B_{0}[p^2, m_{\tilde{g}}, m_{\tilde{t}_2}] \right ),
\end{array}
\eqno{(2.7)}
$$

$$
\begin{array} {lll}
\Sigma^{S~\ddag}(p^2) &=& \frac{C_F}{8 \pi^2} g_{s}^2 m_{\tilde{g}} A_{R} B_{R}
        \left ( B_{0}[p^2, m_{\tilde{g}}, m_{\tilde{t}_1}]
         - B_{0}[p^2, m_{\tilde{g}}, m_{\tilde{t}_2}] \right ).
\end{array}
\eqno{(2.8)}
$$

$m_t$ and $m_{\tilde{t}_n}~(n=1,2)$ involved above are the masses of the top
quark and stop mass eigenstates. In the SU(3) group, $C_F = (N_c^2 - 1)/(2 N_c)$.
Imposing the on-shell renormalization conditions given in Ref.\cite{a11},
one can obtain equations for the renormalized self-energy functions:

$$
      m_t \tilde{Re} \hat{\Sigma}^{L}(m_t^2) +
      \tilde{Re} \hat{\Sigma}^{S~\ddag}(m_t^2) = 0
$$
$$
      m_t \tilde{Re} \hat{\Sigma}^{R}(m_t^2) +
      \tilde{Re} \hat{\Sigma}^{S}(m_t^2) = 0
$$
$$
      m_t \tilde{Re} \hat{\Sigma}^{L}(m_t^2) +
      \tilde{Re} \hat{\Sigma}^{S}(m_t^2) = 0
$$
$$
      m_t \tilde{Re} \hat{\Sigma}^{R}(m_t^2) +
      \tilde{Re} \hat{\Sigma}^{S~\ddag}(m_t^2) = 0
$$
$$
\tilde{Re} \hat{\Sigma}^{L}(m_t^2) + \tilde{Re} \hat{\Sigma}^{R}(m_t^2) +
2 m_t \frac{\partial}{\partial p^2} \tilde{Re} (m_t \hat{\Sigma}^{L}(p^2) +
                                                m_t \hat{\Sigma}^{R}(p^2) +
                                                \hat{\Sigma}^{S}(p^2) +
                                                \hat{\Sigma}^{S~\ddag}(p^2))
                                                |_{p^2=m_{t}^2} = 0
\eqno{(2.9)}
$$

where $\tilde{Re}$ only takes the real part of the loop integral functions
appearing in the self energies. The relevant self-energy and vertex
counterterms are

$$
\begin{array} {lll}
i \delta \Sigma &=& i [
  C_{L} \rlap/p P_{L} + C_{R} \rlap/p P_{R} - C^{-}_{S} P_{L} - C^{+}_{S} P_{R}
  ], \\
  \\
i \delta \Lambda^{\mu} &=& -i e Q_t \gamma^{\mu} [
  C^{-} P_{L} + C^{+} P_{R} ]. \\
\end{array}
\eqno{(2.10)}
$$

where $Q_t$ is the charge of the top quark and

$$
\begin{array} {lll}
C_L &=& C^{-} = \frac{1}{2} (\delta Z^L + \delta Z^{L\dag}), \\
C_R &=& C^{+} = \frac{1}{2} (\delta Z^R + \delta Z^{R\dag}), \\
C^{-}_{S} &=& \frac{m_{t}}{2} (\delta Z^L + \delta Z^{R\dag}) +
              \delta m, \\
C^{+}_{S} &=& \frac{m_{t}}{2} (\delta Z^R + \delta Z^{L\dag}) +
              \delta m.
\end{array}
\eqno{(2.11)}
$$

We choose the field renormalization constant $\delta Z^{R}$ to be real.
Using Eq. (2.9)-(2.11), we obtain the renormalization constants as

$$
\begin{array} {lll}
\delta m &=& \frac{1}{2}
        \left ( m_t \tilde{Re} \Sigma^{L}(m_{t}^{2}) +
                m_t \tilde{Re} \Sigma^{R}(m_{t}^{2}) +
                \tilde{Re} \Sigma^{S}(m_{t}^{2}) +
                \tilde{Re} \Sigma^{S~\ddag}(m_{t}^{2}) \right ),
\end{array}
\eqno{(2.12)}
$$
$$
\begin{array} {lll}
\delta Z^{L} &=& -\Sigma^{L}(m_{t}^{2}) -
    \frac{1}{m_t} \left [ \tilde{Re} \Sigma^{S~\ddag}(m_t^2) -
                          \tilde{Re} \Sigma^{S}(m_t^2) \right ] \\
    &-& m_{t} \frac{\partial}{\partial p^2}
        \left [ m_t \tilde{Re} \Sigma^{L}(p^2) +
                m_t \tilde{Re} \Sigma^{R}(p^2) +
                \tilde{Re} \Sigma^{S}(p^2) +
                \tilde{Re} \Sigma^{S~\ddag}(p^2) \right ] |_{p^2=m_{t}^2}, \\
\end{array}
\eqno{(2.13)}
$$
$$
\begin{array} {lll}
\delta Z^{R} &=& -\Sigma^{R}(m_{t}^{2}) -
    m_{t} \frac{\partial}{\partial p^2}
        \left [ m_t \tilde{Re} \Sigma^{L}(p^2) +
                m_t \tilde{Re} \Sigma^{R}(p^2) +
                \tilde{Re} \Sigma^{S}(p^2) +
                \tilde{Re} \Sigma^{S~\ddag}(p^2) \right ] |_{p^2=m_{t}^2}, \\
\end{array}
\eqno{(2.14)}
$$

  Including all next-to-leading order supersymmetric QCD corrections, the
renormalized amplitude for $t\bar{t}$ pair production in $\gamma \gamma$
collisions is shown as

$$
\begin{array} {lll}
 M_{ren}(\lambda_2, \lambda_1) &=& M_{0}(\lambda_2 ,\lambda_1) +
          \delta M^{1-loop}(\lambda_2, \lambda_1)\\
     &=& M_{0}(\lambda_2, \lambda_1)+\delta{M}^{self}(\lambda_2, \lambda_1) +
          \delta{M}^{vertex}(\lambda_2, \lambda_1) +\\
     &~~& \delta{M}^{box}(\lambda_2, \lambda_1) +
          \delta{M}^{tr}(\lambda_2, \lambda_1)
\end{array}
\eqno{(2.15)}
$$

where $M_{0}$ is the amplitude at the tree level, $\delta M^{self}$,
$\delta M^{vertex}$, $\delta M^{box}$ and $\delta M^{tr}$ represent
the renormalized amplitudes with the next-to-leading order supersymmetric
QCD corrections arising from the self-energy, vertex, box and triangle
diagrams, respectively. Their explicit forms are given by

$$
\begin{array} {lll}
M_{0}(\lambda_2, \lambda_1)= &-&i \frac{e^2 Q_t^2}{\hat{t}-m_t^2}
        \epsilon_{\mu}(p_4,\lambda_2)\epsilon_{\nu}(p_3,\lambda_1)
        \bar{u} (p_2)~\gamma^{\mu}~
        (\rlap/p_2-\rlap/p_4+m_t)\gamma^{\nu}~v(p_1) \\
    &+& ~(p_3 \leftrightarrow p_4, \mu \leftrightarrow \nu,
         \hat{t} \rightarrow \hat{u})
\end{array}
\eqno{(2.16)}
$$

$$
\delta M^{s}(\lambda_2, \lambda_1) = i \frac{e^{2} Q_{t}^{2}}
     {(\hat{t} - m_{t}^{2})^{2}}
   \epsilon_{\mu}(p_4,\lambda_2)\epsilon_{\nu}(p_3,\lambda_1)
       \sum_{N=L,R} \bar{u}_{N} (p_2) \hskip 25mm
$$
$$
\begin{array} {lll}
   &~&~(f_{1,N}^{s(\hat{t})}~\gamma^{\mu}
        \gamma^{\nu} +f_{2,N}^{s(\hat{t})}~p_{2}^{\mu}\gamma^{\nu}+
        f_{3,N}^{s(\hat{t})}~\rlap/p_4 \gamma^{\mu}\gamma^{\nu})~v(p_1) \\
   & & +~(p_3 \leftrightarrow p_4, \mu \leftrightarrow \nu,
        \hat{t} \rightarrow \hat{u})
\end{array}
\eqno{(2.17)}
$$

$$
\delta M^{v}(\lambda_2, \lambda_1) = -i \frac{e^{2} Q_{t}^{2}}{\hat{t} - m_{t}^{2}}
   \epsilon_{\mu}(p_4,\lambda_2)\epsilon_{\nu}(p_3,\lambda_1)~ {\sum_{N=L,R}}
       \bar{u}_{N} (p_2)  \hskip 30mm
$$
$$
\begin{array} {lll}
   &~& ~(f_{1,N}^{v(\hat{t})}~\gamma^{\mu}\gamma^{\nu}
       +f_{2,N}^{v(\hat{t})}~p_{1}^{\nu}~\gamma^{\mu}
       +f_{3,N}^{v(\hat{t})}~p_{2}^{\mu}~\gamma^{\nu} \\
   &~& ~+f_{4,N}^{v(\hat{t})}~p_{1}^{\nu}~p_{2}^{\mu}
       +f_{5,N}^{v(\hat{t})}~\rlap/p_4 \gamma^{\mu}\gamma^{\nu}
       +f_{6,N}^{v(\hat{t})}~\rlap/p_4 p_{1}^{\nu}\gamma^{\mu}
       +f_{7,N}^{v(\hat{t})}~\rlap/p_4 p_{2}^{\mu}\gamma^{\nu})~v(p_1) \\
   & & +~(p_3 \leftrightarrow p_4, \mu \leftrightarrow \nu,
        \hat{t} \rightarrow \hat{u})
\end{array}
\eqno{(2.18)}
$$

$$
\delta M^{b}(\lambda_2, \lambda_1) = i \frac{2 e^2 Q_t^2 g_s^2}
   {\pi^2} \epsilon_{\mu} (p_4,\lambda_{1}) \epsilon_{\nu}
   (p_3,\lambda_{2}) \sum_{N=L,R}^{} \bar{u}_{N} (p_2)  \hskip 25mm
$$
$$
\begin{array} {lll}
   &~& ~\left[ f_{1,N}^{b(\hat{t})}\gamma^{\mu}\gamma^{\nu}
       +f_{2,N}^{b(\hat{t})}\gamma^{\nu}\gamma^{\mu}
       +f_{3,N}^{b(\hat{t})}p_1^{\nu}\gamma^{\mu}
       +f_{4,N}^{b(\hat{t})}p_1^{\mu}\gamma^{\nu} \right. \\
   &~& ~+f_{5,N}^{b(\hat{t})}p_2^{\nu}\gamma^{\mu}
       +f_{6,N}^{b(\hat{t})}p_2^{\mu}\gamma^{\nu}
       +f_{7,N}^{b(\hat{t})}p_1^{\mu}p_1^{\nu}
       +f_{8,N}^{b(\hat{t})}p_1^{\mu}p_2^{\nu}
       +f_{9,N}^{b(\hat{t})}p_2^{\mu}p_1^{\nu} \\
   &~& ~+f_{10,N}^{b(\hat{t})}p_2^{\mu}p_2^{\nu}
       +f_{11,N}^{b(\hat{t})}{\rlap/p_4}\gamma^{\mu}\gamma^{\nu}
       +f_{12,N}^{b(\hat{t})}{\rlap/p_4}\gamma^{\nu}\gamma^{\mu}
       +f_{13,N}^{b(\hat{t})}{\rlap/p_4}p_1^{\mu}p_1^{\nu} \\
   &~& ~+f_{14,N}^{b(\hat{t})}{\rlap/p_4}p_1^{\mu}p_2^{\nu} \left.
       +f_{15,N}^{b(\hat{t})}{\rlap/p_4}p_2^{\mu}p_1^{\nu}
       +f_{16,N}^{b(\hat{t})}{\rlap/p_4}p_2^{\mu}p_2^{\nu} \right ]~v(p_1) \\
   & & +~(p_3 \leftrightarrow p_4, \mu \leftrightarrow \nu,
        \hat{t} \rightarrow \hat{u})
\end{array}
\eqno{(2.19)}
$$

$$
\delta M^{tr}(\lambda_2, \lambda_1)= -i\frac{e^2 Q_t^2 g_s^2}{\pi^2}  g^{\mu \nu}
   \epsilon_{\mu}(p_4,\lambda_2)\epsilon_{\nu}(p_3,\lambda_1)
    {\sum_{N=L,R}} f^{tr}_{N} \bar{u}_{N}(p_2) v(p_1)
\eqno{(2.20)}
$$

Here $\hat{t}=(p_4-p_2)^2$, $\hat{u}=(p_1-p_4)^2 $. The explicit form factors
$f_{i,N}^{s(\hat{t})},f_{i,N}^{v(\hat{t})},f_{i,N}^{b(\hat{t})},f^{tr}_{N}$
are presented in the appendix. The analytical deduction for the renormalized
amplitudes in Eq.(2.15) clearly states the complete cancellation of
ultraviolet divergences, which is required to ensure the correctness
of our calculation.

  The cross section of the process for the polarized photons is given by

$$
 \hat{\sigma}^{\lambda_2,\lambda_1}(\hat{s})=\frac{N_c}{16 \pi \hat{s}
             (\hat{s}-4 m_{t}^2)}
             \int_{\hat{t}^{-}}^{\hat{t}^{+}} d\hat{t} {\sum_{spins}^{}}
             \vert M_{ren}^{\lambda_2 \lambda_1}(\hat{s},\hat{t}) \vert^2,
\eqno{(2.21)}
$$

where $N_c$ is the number of colors, $~\hat{t}^\pm=(m_t^2-\frac{1}{2}\hat{s})
\pm\frac{1}{2}\hat{s}\beta_t~ $ , $ \beta_t=\sqrt{1-4m_t^2/\hat{s}} $. Note
that the summing over the spins is performed only over the final quark pair:

$$
 {\sum_{spins}^{}}\vert M_{ren}^{\lambda_2, \lambda_1}(\hat{s},\hat{t}) \vert^2=
        {\sum_{spins}^{}} \vert
        M_{0}^{\lambda_2, \lambda_1}\vert^2 + 2
        Re \left( {\sum_{spins}^{}} M_{0}^{\lambda_2 \lambda_1 \dag} \cdot
        \delta M_{1-loop}^{\lambda_2, \lambda_1}
        \right).
\eqno{(2.22)}
$$
For polarized massless vector particles we have

$$
 \epsilon^{\mu}_{\lambda_2}(k) \epsilon^{\nu \ast}_{\lambda_1}(k) =
        \frac{\delta_{\lambda_2,\lambda_1}}{2} (-g^{\mu \nu} +
        \frac{p^{\mu}k^{\nu} + p^{\nu}k^{\mu}}{(p \cdot k)} +
        i \lambda_2 \epsilon^{\sigma \mu \rho \nu}
          \frac{k_{\sigma} p_{\rho}}{(p \cdot k)})
\eqno{(2.23)}
$$

where p is an arbitrary light-like Lorentz vector.

The cross section with unpolarized photons is

$$
     \sigma = \frac{1}{4}\sum_{\lambda_1,\lambda_2=-1}^{+1}
              \sigma(\lambda_1,\lambda_2). \\
\eqno{(2.24)}
$$

\begin{flushleft} {\bf III. Numerical results and discussion. } \end{flushleft}
In the numerical evaluation, we take the top quark mass to be $m_t=175~GeV$,
$~\alpha = 1/137.036$ and use the two-loop running coupling constant
$\alpha_s$. For the MSSM parameters, we assume that the supersymmetric weak
eigenstate partners $\tilde{t}_L$ and $\tilde{t}_R$ mix equivalently, namely
the mixing angle $\theta_{\tilde t}=45^{\circ}$. For the other two independent
parameters in stop quark mass matrix, we use the masses of the stop mass
eigenstates, and assume $m_{\tilde{t}_1} \le m_{\tilde{t}_2}$. Current
experiments \cite{a13} constrain the light stop mass eigenstate to be heavier
than 50 GeV. To describe CP violation in $t\bar{t}$ production in polarized
$\gamma\gamma$ fusion, a CP-violating parameter $\xi_{CP}$ is defined as below:

$$
\xi_{CP} = \frac{\sigma(++) - \sigma(--)}{\sigma(++) + \sigma(--)}
\eqno{(3.1)}
$$

We set the CP-violating phase to be $\phi=45^{\circ}$ when evaluating the
cross section in the case of non-zero phase angle, so that the CP-violating
effect can be maximized. \\

In Fig.2(a)(b) the corrections for the cross sections $\sigma(\pm\pm)$ and
$\sigma(\pm\mp)$ for different initial photon linear polarizations are plotted
as functions of the c.m. energy $\sqrt{\hat{s}}$, assuming the following mass
values for supersymmetric particles involved:

$$
m_{\tilde{g}}=100~GeV,~~m_{\tilde{t}_1}=100~GeV,~~m_{\tilde{t}_2}= 450~GeV
\eqno{(3.2)}
$$

  Unlike the corrections from the SM \cite{a5}, all the contributions of the
MSSM are negative in all the polarization cases. The relative correction
$\delta$, defined as $(\hat{\sigma}-\hat{\sigma}_{0})/\hat{\sigma}_{0}$,
is generally in the order of $10^{-2}$ which is less than that of the
Standard Model QCD corrections \cite{a5}. But it can also reach about
$10 \%$ in some points. Fig.2(a) and Fig.2(b) show that the corrections
will be suppressed in the case of introducing a non-zero CP-violating
phase angle for both of the polarized photon fusion modes under the
conditions of Eq.(3.2). The dependence of the CP-violation parameter
$\xi_{CP}$ on $\sqrt{\hat{s}}$ is given in Fig.2(c). It shows that in
the c.m. energy range around the $t \bar{t}$ threshold, CP is strongly
violated, where $\xi_{CP}$ can be above $2.6 \%$. The curve drops sharply
with increasing $\sqrt{\hat{s}}$. From the definition of the CP-violating
parameter $\xi_{CP}$ given in Eq.(3.1), one can understand that the feature
of the curve in Fig.2(c) is simply the result from Fig.2(a,b).

Since the present experimental results can not exclude the existence of very
light gluinos and there has been renewed interest in this case recently, we
also calculate the relative corrections to the cross sections with a very light
gluino mass, as functions of $\sqrt{\hat{s}}$ when

$$
m_{\tilde{g}}=5~GeV,~~m_{\tilde{t}_1}=50~GeV,~~m_{\tilde{t}_2}=250~GeV
\eqno{(3.3)}
$$

The relative corrections $\delta(\pm,\pm)$ and $\delta(\pm,\mp)$
for different photon linear polarizations are plotted in Fig.3(a)and Fig.3(b)
respectively. The curves go from positive to negative
values as $\sqrt{\hat{s}}$ increases. The relative corrections are positive
around $t \bar{t}$ threshold, whereas they are always negative with the
conditions Eq.(3.2) where the gluino mass is much heavier.
Although in most energy regions the absolute corrections are small compared
with the results given in Fig.2(a,b), the maximal value of relative
correction is still about $3.5 \%$ around $\sqrt{\hat{s}} = 500~GeV$ which can
be seen in Fig.3(a). The high peaks shown in Fig.3(a) are the results from
resonance effect when $\sqrt{\hat{s}} \sim 2 m_{\tilde{t}_2} = 500~GeV$,
which comes from the contributions of the triangle diagrams in Fig.1(i).
The CP violation quantities in this case are plotted in Fig.3(c). All the
values of $\xi_{CP}$ have negative signs, and are about an order of magnitude
lower than the values in Fig.2(c), varying from $-0.184 \%$ to $-0.296 \%$.

As a check, we also calculate the corrections with extending c.m. energy
$\sqrt{\hat{s}}$ up to the region between $2 \sim 5~TeV$ and taking the same
parameters of Eq.(3.2) and Eq.(3.3) respectively. The absolute corrections
get smaller with increasing $\sqrt{\hat{s}}$. Noteworthily the convergency
of the correction with heavy gluino and squarks is much faster than that
with light gluino and squarks. It proves that the decoupling really takes
place at high energies, and the results of our cross section spectra at
one-loop level satisfy the unitarity which is expected on theoretical grounds.

According to Eq.(2.24), we can get the relationship between the unpolarized
cross section and the gluino mass, which is shown in Fig.4(a) with

$$
m_{\tilde{t}_1}=100~GeV,~~m_{\tilde{t}_2}=450~GeV~~and~~\sqrt{\hat{s}}=500~GeV
\eqno{(3.4)}
$$

From the figure, we can see that the corrections jump abruptly at the point
$m_{\tilde{g}} = 75~GeV$. This sudden change is due to the influence of a
singularity in top quark wave function renormalization at the point
$m_t = m_{\tilde{g}} + m_{\tilde{t}_1}$.
This singularity originates from the renormalization constants $\delta Z^{L}$
and $\delta Z^{R}$ of the top quark wave function shown in Eqs.(2.13) and
(2.14). The effect of the singularity makes the absolute value of the
correction get a large increment near the singularity point.
However at the gluino mass range which is far away from the singularity
mass region, the correction approaches to a small value quickly. This
implies that there exists decoupling with heavy gluino in the SUSY QCD
corrections. The corresponding prediction for the CP-violation parameter
$\xi_{CP}$ as a function of $m_{\tilde{g}}$ is shown in Fig.4(b). There the
singularity effect from the top quark wave function renormalization exists
at the point $m_{\tilde{g}} = 75~GeV$.
The influence of the singularity also can be seen from the
fact that in Fig.3(a)(b) the corrections are distinctly much smaller than
those in Fig.2(a)(b). That is just because the parameters for Fig.3(a)(b)
as shown in Eq.(3.3) are far away from the singularity point.

The relations between the corrections of the unpolarized cross section
and the two stop quark masses $m_{\tilde{t}_1}$ and $m_{\tilde{t}_2}$
are depicted in Fig.5(a) and Fig.6(a) respectively.
In Fig.5(a), the large corrections near the point $m_{\tilde{t}_1} = 75~GeV$
are again due to the influence of the singularity.
However, the corrections show the decoupling with the masses of stop quarks,
when $m_{\tilde{t}_i}~(i=1,2)$ are far away from the singularity points
$m_{\tilde{t}_i}=m_t - m_{\tilde{g}}$.
The decoupling effect can be seen more clearly in the heavy stop case.
In Fig.6(a) the curves of corrections are rather flat in the heavy
$\tilde{t}_2$ region, and the small waves around $m_{\tilde{t}_2} = 250~GeV$
are merely due to the resonance effect when
$\sqrt{\hat{s}} = 2 m_{\tilde{t}_2}$.
The corresponding dependence of parameter $\xi_{CP}$ on these two stop
quark masses is given in Fig.5(b),6(b).

Fig.4(a), Fig.5(a) and Fig.6(a) also show clearly that the absolute
corrections for $\phi=45^{\circ}$ are apparently smaller than those
for $\phi=0^{\circ}$ in unpolarized photon fusion,
except in the light gluino mass region in Fig.4(a) which is simply
due to the singularity effect in top quark wave function renormalization.
The fact that the corrections are
always maximal for vanishing CP violation is simply due to the fact that
with introducing a non-zero CP-violating phase angle in the gluino-stop-top
coupling, the real part of this coupling strength will be reduced down
quantitatively. Further analyses show that this reduction will lead a
decrease in the absolute values of the real parts of form factors,
whereas the imaginary parts of form factors do not contribute to the
cross section. But around $m_{\tilde{t}_1} \sim \sqrt{\hat{s}}/2$
mass region shown in Fig.5(a), the situation is more complicated and
this feature does not exist.

\vskip 5mm
\begin{flushleft} {\bf IV. Conclusion} \end{flushleft}
  In this work we have studied the one-loop supersymmetric QCD corrections of
the process $\gamma(\lambda_2) \gamma(\lambda_1) \rightarrow t\bar{t}$.

  From the results of numerical calculation, we can conclude that the
singularity of the top quark wave function renormalization affects the
corrections heavily. Large corrections can be expected near the singularity
region which satisfies $m_t=m_{\tilde{g}}+m_{\tilde{t}_i}~~(i=1,2)$. This
effect is shown in Fig.4(a), 5(a) and 6(a), where the large corrections are
all due to the singularity effect. This singularity effect can be also used
to explain why the corrections are large in Fig.2(a), whereas the corrections
in Fig.3(a) are small where the parameters are far away from all the
singulatriy points. However the corrections are insensitive to the masses of
supersymmetric particles $\tilde{t}_i~(i=1,2)$ and $\tilde{g}$ when their
masses are heavy and far away from singularity points. This feature is just
the manifestation of the decoupling effects.

  In any case, with the fusion of photons polarized with parallel spin
directions, the corrections are always significant near the $t \bar{t}$
threshold. Furthermore, if the CP-violating phase really exists in the
squark mixing matrix or in the Majorana mass term of the gluino predicted
by the MSSM, the CP-violating parameter could be expected to be the order
of $10^{-3} \sim 10^{-2}$. Therefore testing CP-violation in this process
is one of the promising tasks for a future photon-photon collider.

This work was supported in part by the National Natural Science Foundation
of China and the National Committee of Science and Technology of China.
Part of this work was done when two of the authers, Ma Wen-Gan and Yu Zeng-Hui,
visited the University Vienna under the exchange agreement
(project number: IV.B.12).

\newpage
\vspace{0.2in}
        \begin{center} {\Large Appendix}\end{center}
\vspace{0.1in}

    The form factors $f_{i,N}^{s(\hat{t})}$ can be expressed as
$$
\begin{array} {lll}
f_{1,R}^{s(\hat{t})} &=&
        \frac{g_{s}^2 C_F}{4 \pi^2} (p_{2} \cdot p_{4}) \left\{
          A_{L} B_{R} m_{t} B_{1}[-p_{2} + p_{4}, m_{\tilde{g}}, m_{\tilde{t}_1}]
          + (L \leftrightarrow R, m_{\tilde{t}_1} \rightarrow m_{\tilde{t}_2})
          \right\} \\
    &-& \frac{g_{s}^2 C_F}{4 \pi^2} (p_{2} \cdot p_{4}) A_{L} B_{L} \left\{
          m_{\tilde{g}} B_{0}[-p_{2} + p_{4}, m_{\tilde{g}}, m_{\tilde{t}_1}]
          + (m_{\tilde{g}} \rightarrow -m_{\tilde{g}}, m_{\tilde{t}_1} \rightarrow m_{\tilde{t}_2})
          \right\} \\
    &+& 2 (p_{2} \cdot p_{4}) (C_{S}^{-} - m_{t} C_{R}) \\
\end{array}
\eqno{(A1)}
$$

$$
\begin{array} {lll}
f_{2,R}^{s(\hat{t})} &=&
        \frac{g_{s}^2 C_F}{4 \pi^2}  \left\{
          (2 A_{L} B_{R} ~p_{2} \cdot p_{4} - m_t^2) B_{1}
          [-p_{2} + p_{4}, m_{\tilde{g}}, m_{\tilde{t}_1}]
          + (L \leftrightarrow R, m_{\tilde{t}_1} \rightarrow m_{\tilde{t}_2})
          \right\} \\
    &+& \frac{g_{s}^2 C_F}{4 \pi^2} (A_{L} B_{L} + A_{R} B_{R}) m_{t} \left\{
          m_{\tilde{g}} B_{0}
          [-p_{2} + p_{4}, m_{\tilde{g}}, m_{\tilde{t}_1}]
          + (m_{\tilde{g}} \rightarrow -m_{\tilde{g}}, m_{\tilde{t}_1} \rightarrow m_{\tilde{t}_2})
          \right\} \\
    &-& 2 m_{t} (C_{S}^{-} + C_{S}^{+})
          + 2 m_{t}^{2} (C_{L} + C_{R}) - 4~(p_{2} \cdot p_{4})~C_{R} \\
\end{array}
\eqno{(A2)}
$$

$$
\begin{array} {lll}
f_{3,R}^{s(\hat{t})} &=& \frac{1}{2} f_{2,R}^{s(\hat{t})}
\end{array}
\eqno{(A3)}
$$

$$
\begin{array} {lll}
f_{i,L}^{s(\hat{t})} &=& f_{i,R}^{s(\hat{t})}
        (R \leftrightarrow L, C_{S}^{+} \leftrightarrow C_{S}^{-}) \\
        \\
\end{array}
\eqno{(A4)}
$$
where the definitions of the renormalization quantities $C_{S}^{\pm}$ and
$C_{R,L}$ can be found in Eq.(2.11), and those of $A_{R,L}$ and $B_{R,L}$
in Eq.(2.3). The replacement $m_{\tilde{g}} \rightarrow -m_{\tilde{g}}$
is not performed on the argument of loop integral functions $B$, $C$ and $D$.

    The form factors of triangle, vertex and box diagrams are given
as follows
$$
\begin{array} {lll}
f^{tr}_{R} &=& \frac{m_t C_F}{4} \left\{
        (A_{R} B_{L} C_{11} - A_{R} B_{L} C_{12} + A_{L} B_{R} C_{12})
          [-p_{2}, p_{1} + p_{2}, m_{\tilde{g}}, m_{\tilde{t}_1}, m_{\tilde{t}_1}] \right. \\
      &~& \left. + (L \leftrightarrow R, m_{\tilde{t}_1} \rightarrow m_{\tilde{t}_2})
          \right\} \\
    &-& \frac{C_F}{4} A_{L} B_{L} \left\{ m_{\tilde{g}} C_{0}
          [-p_{2}, p_{1} + p_{2}, m_{\tilde{g}}, m_{\tilde{t}_1}, m_{\tilde{t}_1}] \right. \\
      &~& \left. ~+ (m_{\tilde{g}} \rightarrow -m_{\tilde{g}}, m_{\tilde{t}_1} \rightarrow m_{\tilde{t}_2})
          \right\} \\
\end{array}
\eqno{(A5)}
$$

$$
\begin{array} {lll}
f^{tr}_{L} &=& f_{R}^{tr} (R \leftrightarrow L) \\
\end{array}
\eqno{(A6)}
$$

$$
\begin{array} {lll}
f_{1,R}^{v(\hat{t})} &=&
        \frac{g_{s}^2 C_F}{4 \pi^2} m_{t} \left\{
          (A_{L} B_{R} - A_{R} B_{L}) C_{24}
          [-p_{2}, p_{4}, m_{\tilde{g}}, m_{\tilde{t}_1}, m_{\tilde{t}_1}]
          + (L \leftrightarrow R, m_{\tilde{t}_1} \rightarrow m_{\tilde{t}_2})
          \right\} \\
    &+& m_{t} (C_{R} - C_{L}) \\
\end{array}
\eqno{(A7)}
$$

$$
\begin{array} {lll}
f_{2,R}^{v(\hat{t})} &=&
       -\frac{g_{s}^2 C_F}{4 \pi^2}  2 (p_{1} \cdot p_{3}) \left\{
          A_{L} B_{R} (C_{12} + C_{23})
          [p_{1}, -p_{3}, m_{\tilde{g}}, m_{\tilde{t}_1}, m_{\tilde{t}_1}]
          + (L \leftrightarrow R, m_{\tilde{t}_1} \rightarrow m_{\tilde{t}_2})
          \right\} \\
\end{array}
\eqno{(A8)}
$$

$$
\begin{array} {lll}
f_{3,R}^{v(\hat{t})} &=&
        \frac{g_{s}^2 C_F}{4 \pi^2} \left\{
          (4 A_{L} B_{R} C_{24} - m_{t}^2 (C_{11} + C_{21})
          + 2~p_{2} \cdot p_{4}~A_{L} B_{R} (C_{12} + C_{23}))
          [-p_{2}, p_{4}, m_{\tilde{g}}, m_{\tilde{t}_1}, m_{\tilde{t}_1}] + \right. \\
      &~& \left. (L \leftrightarrow R, m_{\tilde{t}_1} \rightarrow m_{\tilde{t}_2})
          \right\} \\
    &+& \frac{g_{s}^2 C_F}{4 \pi^2} m_{t} (A_{L} B_{L} + A_{R} B_{R}) \left\{
          m_{\tilde{g}} (C_{0} + C_{11})
          [-p_{2}, p_{4}, m_{\tilde{g}}, m_{\tilde{t}_1}, m_{\tilde{t}_1}]
          + (m_{\tilde{g}} \rightarrow -m_{\tilde{g}}, m_{\tilde{t}_1} \rightarrow m_{\tilde{t}_2})
          \right\} \\
    &+& 4 C_{R} \\
\end{array}
\eqno{(A9)}
$$

$$
\begin{array} {lll}
f_{4,R}^{v(\hat{t})} &=&
        \frac{g_{s}^2 C_F}{2 \pi^2} m_{t} \left\{
          (A_{L} B_{R} C_{11} + A_{R} B_{L} C_{12} - A_{L} B_{R} C_{12} + A_{L} B_{R} C_{21}
          + \right. \\
      &~& \left. A_{R} B_{L} C_{23} - A_{L} B_{R} C_{23})
          [p_{1}, -p_{3}, m_{\tilde{g}}, m_{\tilde{t}_1}, m_{\tilde{t}_1}]
          + (L \leftrightarrow R, m_{\tilde{t}_1} \rightarrow m_{\tilde{t}_2})
          \right\} \\
    &-& \frac{g_{s}^2 C_F}{2 \pi^2} A_{L} B_{L} \left\{
          m_{\tilde{g}} (C_{0} + C_{11})
          [p_{1}, -p_{3}, m_{\tilde{g}}, m_{\tilde{t}_1}, m_{\tilde{t}_1}]
          + (m_{\tilde{g}} \rightarrow -m_{\tilde{g}}, m_{\tilde{t}_1} \rightarrow m_{\tilde{t}_2})
          \right\} \\
\end{array}
\eqno{(A10)}
$$

$$
\begin{array} {lll}
f_{5,R}^{v(\hat{t})} &=&
        \frac{g_{s}^2 C_F}{2 \pi^2} \left\{
          A_{L} B_{R} C_{24}[-p_{2}, p_{4}, m_{\tilde{g}}, m_{\tilde{t}_1}, m_{\tilde{t}_1}]
          + (L \leftrightarrow R, m_{\tilde{t}_1} \rightarrow m_{\tilde{t}_2})
          \right\} \\
    &+& 2 C_{R} \\
\end{array}
\eqno{(A11)}
$$

$$
\begin{array} {lll}
f_{6,R}^{v(\hat{t})} &=& \frac{1}{2} f_{4,R}^{v(\hat{t})} \\
\end{array}
\eqno{(A12)}
$$

$$
\begin{array} {lll}
f_{7,R}^{v(\hat{t})} &=&
        \frac{g_{s}^2 C_F}{4 \pi^2} m_{t} \left\{
          (A_{R} B_{L} C_{11} - A_{R} B_{L} C_{12} + A_{L} B_{R} C_{12} + A_{R} B_{L} C_{21} -\right. \\
      &~& \left. A_{R} B_{L} C_{23} + A_{L} B_{R} C_{23})
          [-p_{2}, p_{4}, m_{\tilde{g}}, m_{\tilde{t}_1}, m_{\tilde{t}_1}]
          + (L \leftrightarrow R, m_{\tilde{t}_1} \rightarrow m_{\tilde{t}_2})
          \right\} \\
    &-& \frac{g_{s}^2 C_F}{4 \pi^2} A_{L} B_{L} \left\{
          m_{\tilde{g}} (C_{0} + C_{11})
          [-p_{2}, p_{4}, m_{\tilde{g}}, m_{\tilde{t}_1}, m_{\tilde{t}_1}]
          + (m_{\tilde{g}} \rightarrow -m_{\tilde{g}}, m_{\tilde{t}_1} \rightarrow m_{\tilde{t}_2})
          \right\} \\
\end{array}
\eqno{(A13)}
$$

$$
\begin{array} {lll}
f_{i,L}^{v(\hat{t})} &=& f_{i,R}^{v(\hat{t})}
        (R \leftrightarrow L, C_{S}^{+} \leftrightarrow C_{S}^{-}) \\
\end{array}
\eqno{(A14)}
$$

$$
\begin{array} {lll}
f_{1,R}^{b(\hat{t})} &=&
        \frac{C_F}{8} m_{t} \left\{ (A_{R} B_{L} D_{311} - A_{R} B_{L} D_{313} + A_{L} B_{R} D_{313}) \right. \\
      &~& \left. [-p_{2}, p_{4}, p_{3}, m_{\tilde{g}}, m_{\tilde{t}_1}, m_{\tilde{t}_1}, m_{\tilde{t}_1}]
          + (L \leftrightarrow R, m_{\tilde{t}_1} \rightarrow m_{\tilde{t}_2})
          \right\} \\
    &-& \frac{C_F}{8} A_{L} B_{L} \left\{
          m_{\tilde{g}} D_{27}[-p_{2}, p_{4}, p_{3}, m_{\tilde{g}}, m_{\tilde{t}_1}, m_{\tilde{t}_1}, m_{\tilde{t}_1}]
          + (m_{\tilde{g}} \rightarrow -m_{\tilde{g}}, m_{\tilde{t}_1} \rightarrow m_{\tilde{t}_2})
          \right\} \\
\end{array}
\eqno{(A15)}
$$

$$
\begin{array} {lll}
f_{2,R}^{b(\hat{t})} &=& f_{1,R}^{b(\hat{t})} \\
\end{array}
\eqno{(A16)}
$$

$$
\begin{array} {lll}
f_{3,R}^{b(\hat{t})} &=&
        - \frac{C_F}{4} A_{L} B_{R} (D_{27} + D_{312})
          [-p_{2}, p_{4}, p_{3}, m_{\tilde{g}}, m_{\tilde{t}_1}, m_{\tilde{t}_1}, m_{\tilde{t}_1}] \\
    &+& (L \leftrightarrow R, m_{\tilde{t}_1} \rightarrow m_{\tilde{t}_2}) \\
\end{array}
\eqno{(A17)}
$$

$$
\begin{array} {lll}
f_{4,R}^{b(\hat{t})} &=&
        - \frac{C_F}{4} A_{L} B_{R} D_{313}
          [-p_{2}, p_{4}, p_{3}, m_{\tilde{g}}, m_{\tilde{t}_1}, m_{\tilde{t}_1}, m_{\tilde{t}_1}] \\
    &+& (L \leftrightarrow R, m_{\tilde{t}_1} \rightarrow m_{\tilde{t}_2}) \\
\end{array}
\eqno{(A18)}
$$

$$
\begin{array} {lll}
f_{5,R}^{b(\hat{t})} &=&
        \frac{C_F}{4} A_{L} B_{R} (D_{311} - D_{312})
        [-p_{2}, p_{4}, p_{3}, m_{\tilde{g}}, m_{\tilde{t}_1}, m_{\tilde{t}_1}, m_{\tilde{t}_1}] \\
    &+& (L \leftrightarrow R, m_{\tilde{t}_1} \rightarrow m_{\tilde{t}_2}) \\
\end{array}
\eqno{(A19)}
$$

$$
\begin{array} {lll}
f_{6,R}^{b(\hat{t})} &=&
        \frac{C_F}{4} A_{L} B_{R} (D_{27} + D_{311} - D_{313})
        [-p_{2}, p_{4}, p_{3}, m_{\tilde{g}}, m_{\tilde{t}_1}, m_{\tilde{t}_1}, m_{\tilde{t}_1}] \\
    &+& (L \leftrightarrow R, m_{\tilde{t}_1} \rightarrow m_{\tilde{t}_2}) \\
\end{array}
\eqno{(A20)}
$$

$$
\begin{array} {lll}
f_{7,R}^{b(\hat{t})} &=& \frac{C_F}{4}
        m_{t} \left\{ (A_{R} B_{L} D_{23} - A_{L} B_{R} D_{23} - A_{R} B_{L} D_{25} - A_{R} B_{L} D_{310} + \right. \\
      &~& A_{R} B_{L} D_{39} - A_{L} B_{R} D_{39})
          [-p_{2}, p_{4}, p_{3}, m_{\tilde{g}}, m_{\tilde{t}_1}, m_{\tilde{t}_1}, m_{\tilde{t}_1}] \\
      &~& \left.+ (L \leftrightarrow R, m_{\tilde{t}_1} \rightarrow m_{\tilde{t}_2}) \right\} \\
    &+& \frac{C_F}{4} A_{L} B_{L} \left\{ m_{\tilde{g}} (D_{13} + D_{26})
        [-p_{2}, p_{4}, p_{3}, m_{\tilde{g}}, m_{\tilde{t}_1}, m_{\tilde{t}_1}, m_{\tilde{t}_1}] \right. \\
      &~& \left. + (m_{\tilde{g}} \rightarrow -m_{\tilde{g}}, m_{\tilde{t}_1} \rightarrow m_{\tilde{t}_2})
        \right\} \\
\end{array}
\eqno{(A21)}
$$

$$
\begin{array} {lll}
f_{8,R}^{b(\hat{t})} &=& \frac{C_F}{4}
        m_{t} \left\{ (A_{R} B_{L} D_{35} - A_{R} B_{L} D_{310} - A_{R} B_{L} D_{37} + A_{L} B_{R} D_{37} + \right. \\
      &~& A_{R} B_{L} D_{39} - A_{L} B_{R} D_{39})
          [-p_{2}, p_{4}, p_{3}, m_{\tilde{g}}, m_{\tilde{t}_1}, m_{\tilde{t}_1}, m_{\tilde{t}_1}] \\
      &~& \left.+ (L \leftrightarrow R, m_{\tilde{t}_1} \rightarrow m_{\tilde{t}_2}) \right\} \\
    &-& \frac{C_F}{4} A_{L} B_{L} \left\{ m_{\tilde{g}} (D_{25} - D_{26})
        [-p_{2}, p_{4}, p_{3}, m_{\tilde{g}}, m_{\tilde{t}_1}, m_{\tilde{t}_1}, m_{\tilde{t}_1}] \right. \\
      &~& \left. + (m_{\tilde{g}} \rightarrow -m_{\tilde{g}}, m_{\tilde{t}_1} \rightarrow m_{\tilde{t}_2})
        \right\} \\
\end{array}
\eqno{(A22)}
$$

$$
\begin{array} {lll}
f_{9,R}^{b(\hat{t})} &=& \frac{C_F}{4}
        m_{t} \left\{ (A_{R} B_{L} D_{11} - A_{R} B_{L} D_{13} + A_{L} B_{R} D_{13} + A_{R} B_{L} D_{21} + \right. \\
      &~& A_{R} B_{L} D_{23} - A_{L} B_{R} D_{23} + A_{R} B_{L} D_{24} - 2 A_{R} B_{L} D_{25} + \\
      &~& A_{L} B_{R} D_{25} - A_{R} B_{L} D_{26} + A_{L} B_{R} D_{26} - 2 A_{R} B_{L} D_{310} + \\
      &~& A_{L} B_{R} D_{310} + A_{R} B_{L} D_{34} + A_{R} B_{L} D_{39} - A_{L} B_{R} D_{39}) \\
      &~& ~[-p_{2}, p_{4}, p_{3}, m_{\tilde{g}}, m_{\tilde{t}_1}, m_{\tilde{t}_1}, m_{\tilde{t}_1}] \\
      &~& \left.+ (L \leftrightarrow R, m_{\tilde{t}_1} \rightarrow m_{\tilde{t}_2}) \right\} \\
    &-& \frac{C_F}{4} A_{L} B_{L} \left\{ m_{\tilde{g}} (D_{0} + D_{11} + D_{12} - D_{13} + D_{24} - D_{26}) \right. \\
      &~& ~[-p_{2}, p_{4}, p_{3}, m_{\tilde{g}}, m_{\tilde{t}_1}, m_{\tilde{t}_1}, m_{\tilde{t}_1}] \\
      &~& \left. + (m_{\tilde{g}} \rightarrow -m_{\tilde{g}}, m_{\tilde{t}_1} \rightarrow m_{\tilde{t}_2})
        \right\} \\
\end{array}
\eqno{(A23)}
$$

$$
\begin{array} {lll}
f_{10,R}^{b(\hat{t})} &=& \frac{C_F}{4}
        m_{t} \left\{ (A_{R} B_{L} D_{24} - A_{R} B_{L} D_{21} + A_{R} B_{L} D_{25} - A_{L} B_{R} D_{25} - \right. \\
      &~& A_{R} B_{L} D_{26} + A_{L} B_{R} D_{26} - A_{R} B_{L} D_{31} - 2 A_{R} B_{L} D_{310} + \\
      &~& A_{L} B_{R} D_{310} + A_{R} B_{L} D_{34} + 2 A_{R} B_{L} D_{35} - A_{L} B_{R} D_{35} - \\
      &~& A_{R} B_{L} D_{37} + A_{L} B_{R} D_{37} + A_{R} B_{L} D_{39} - A_{L} B_{R} D_{39}) \\
      &~& ~[-p_{2}, p_{4}, p_{3}, m_{\tilde{g}}, m_{\tilde{t}_1}, m_{\tilde{t}_1}, m_{\tilde{t}_1}] \\
      &~& \left.+ (L \leftrightarrow R, m_{\tilde{t}_1} \rightarrow m_{\tilde{t}_2}) \right\} \\
    &+& \frac{C_F}{4} A_{L} B_{L} \left\{ m_{\tilde{g}} (D_{11} - D_{12} + D_{21} - D_{24} - D_{25} + D_{26}) \right. \\
      &~& ~[-p_{2}, p_{4}, p_{3}, m_{\tilde{g}}, m_{\tilde{t}_1}, m_{\tilde{t}_1}, m_{\tilde{t}_1}] \\
      &~& \left. + (m_{\tilde{g}} \rightarrow -m_{\tilde{g}}, m_{\tilde{t}_1} \rightarrow m_{\tilde{t}_2})
        \right\} \\
\end{array}
\eqno{(A24)}
$$

$$
\begin{array} {lll}
f_{11,R}^{b(\hat{t})} &=&
        -\frac{C_F}{8} A_{L} B_{R} (D_{312} - D_{313}) \\
       &~& ~[-p_{2}, p_{4}, p_{3}, m_{\tilde{g}}, m_{\tilde{t}_1}, m_{\tilde{t}_1}, m_{\tilde{t}_1}] \\
     &+& (L \leftrightarrow R, m_{\tilde{t}_1} \rightarrow m_{\tilde{t}_2}) \\
\end{array}
\eqno{(A25)}
$$

$$
\begin{array} {lll}
f_{12,R}^{b(\hat{t})} &=& f_{11,R}^{b(\hat{t})} \\
\end{array}
\eqno{(A26)}
$$

$$
\begin{array} {lll}
f_{13,R}^{b(\hat{t})} &=& \frac{C_F}{4}
        A_{L} B_{R} (D_{26} + D_{38} - D_{39} - D_{23}) \\
     &~& ~[-p_{2}, p_{4}, p_{3}, m_{\tilde{g}}, m_{\tilde{t}_1}, m_{\tilde{t}_1}, m_{\tilde{t}_1}] \\
     &+& (L \leftrightarrow R, m_{\tilde{t}_1} \rightarrow m_{\tilde{t}_2}) \\
\end{array}
\eqno{(A27)}
$$

$$
\begin{array} {lll}
f_{14,R}^{b(\hat{t})} &=& \frac{C_F}{4}
        A_{L} B_{R} (D_{37} + D_{38} - D_{39} - D_{310}) \\
     &~& ~[-p_{2}, p_{4}, p_{3}, m_{\tilde{g}}, m_{\tilde{t}_1}, m_{\tilde{t}_1}, m_{\tilde{t}_1}] \\
     &+& (L \leftrightarrow R, m_{\tilde{t}_1} \rightarrow m_{\tilde{t}_2}) \\
\end{array}
\eqno{(A28)}
$$

$$
\begin{array} {lll}
f_{15,R}^{b(\hat{t})} &=& \frac{C_F}{4}
        A_{L} B_{R} (D_{13} - D_{12} - D_{22} - D_{23} - D_{24} + D_{25} + \\
     &~& 2 D_{26} + D_{310} - D_{36} + D_{38} - D_{39}) \\
     &~& ~[-p_{2}, p_{4}, p_{3}, m_{\tilde{g}}, m_{\tilde{t}_1}, m_{\tilde{t}_1}, m_{\tilde{t}_1}] \\
     &+& (L \leftrightarrow R, m_{\tilde{t}_1} \rightarrow m_{\tilde{t}_2}) \\
\end{array}
\eqno{(A29)}
$$

$$
\begin{array} {lll}
f_{16,R}^{b(\hat{t})} &=& \frac{C_F}{4}
        A_{L} B_{R} (D_{24} - D_{22} - D_{25} + D_{26} + D_{34} - D_{35} - \\
     &~& D_{36} + D_{37} + D_{38} - D_{39}) \\
     &~& ~[-p_{2}, p_{4}, p_{3}, m_{\tilde{g}}, m_{\tilde{t}_1}, m_{\tilde{t}_1}, m_{\tilde{t}_1}] \\
     &+& (L \leftrightarrow R, m_{\tilde{t}_1} \rightarrow m_{\tilde{t}_2}) \\
\end{array}
\eqno{(A30)}
$$

$$
\begin{array} {lll}
f_{i,L}^{b(\hat{t})} = f_{i,R}^{b(\hat{t})} (R \leftrightarrow L) \\
\end{array}
\eqno{(A31)}
$$

where
$$
(p_3-p_1)^2=\hat{t},~~\hat{s}=(p_1+p_2)^2,
$$
$$
2 p_1\cdot p_3=2 p_2\cdot p_4=m_t^2-\hat{t},
$$
$$
2 p_3\cdot p_4=\hat{s},~~2 p_1\cdot p_4=2 p_2\cdot p_3=m_t^2-\hat{u}
$$
$$
2 p_1\cdot p_2=\hat{s}-2 m_t^2,~~\hat{s}+\hat{t}+\hat{u}=2 m_t^2.
$$

  All the definitions of loop integral functions $A$, $B$, $C$ and $D$ used in
our paper can be found in Ref.\cite{a14}. The numerical calculation of the
vector and tensor loop integral functions can be traced back to the four
scalar loop integrals $A_0$, $B_0$, $C_0$ and $D_0$ in Ref.\cite{a15}.

\vfil
\eject

\begin{center}
{\large Figure Captions}
\end{center}

{\bf Fig.1} Feynman diagrams for contributions from
        the tree-level and next-to-leading order in Supersymmetric
        QCD for the $\gamma \gamma \rightarrow t \bar{t}$ process:
(a-b) tree level diagrams; (c-e) self-energy diagrams;
(f-g) vertex diagrams; (h) box diagrams: (i) triangle diagrams.
The dashed lines represent $\tilde{t}_1,\tilde{t}_2$ for (c-i).

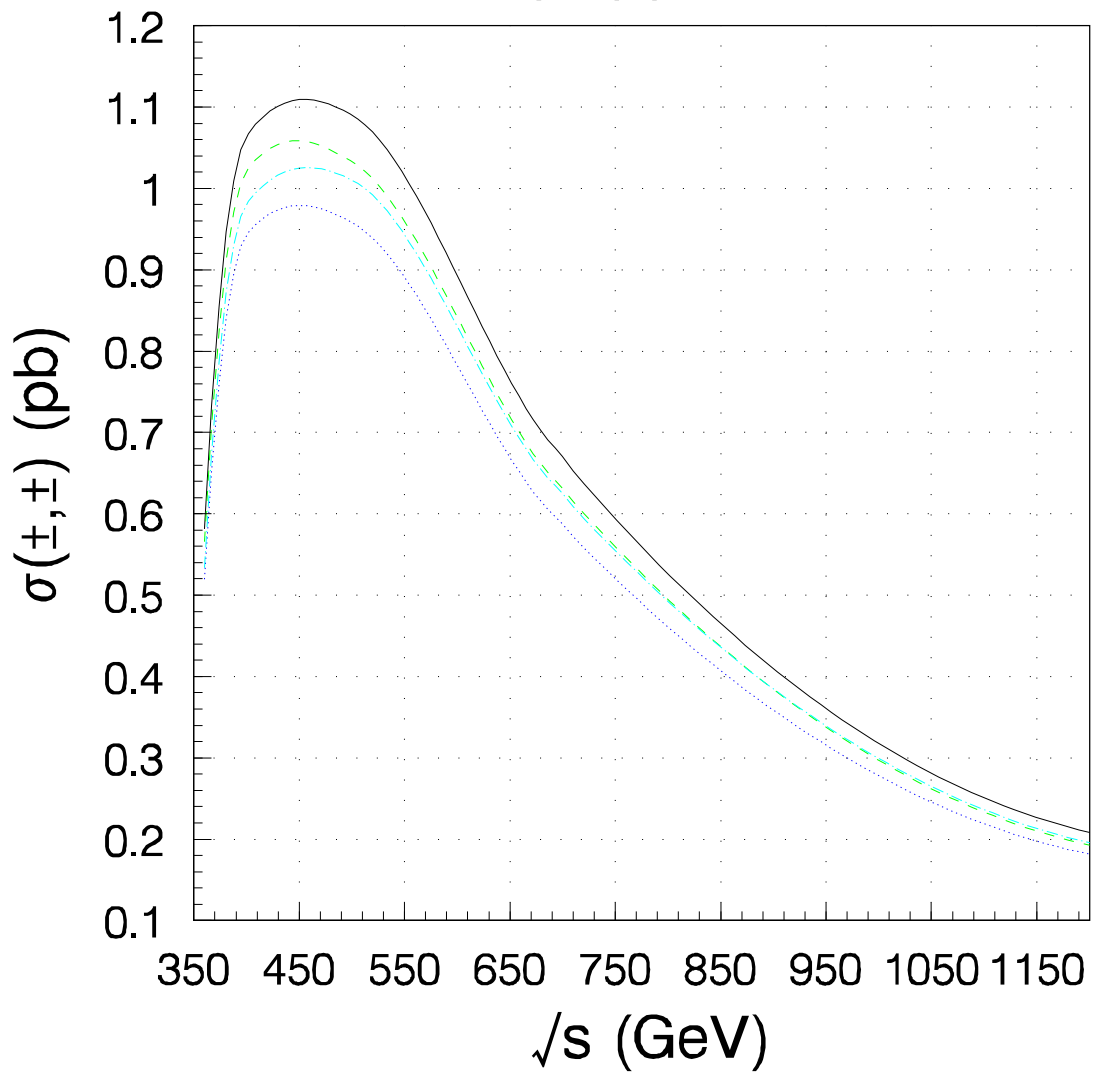
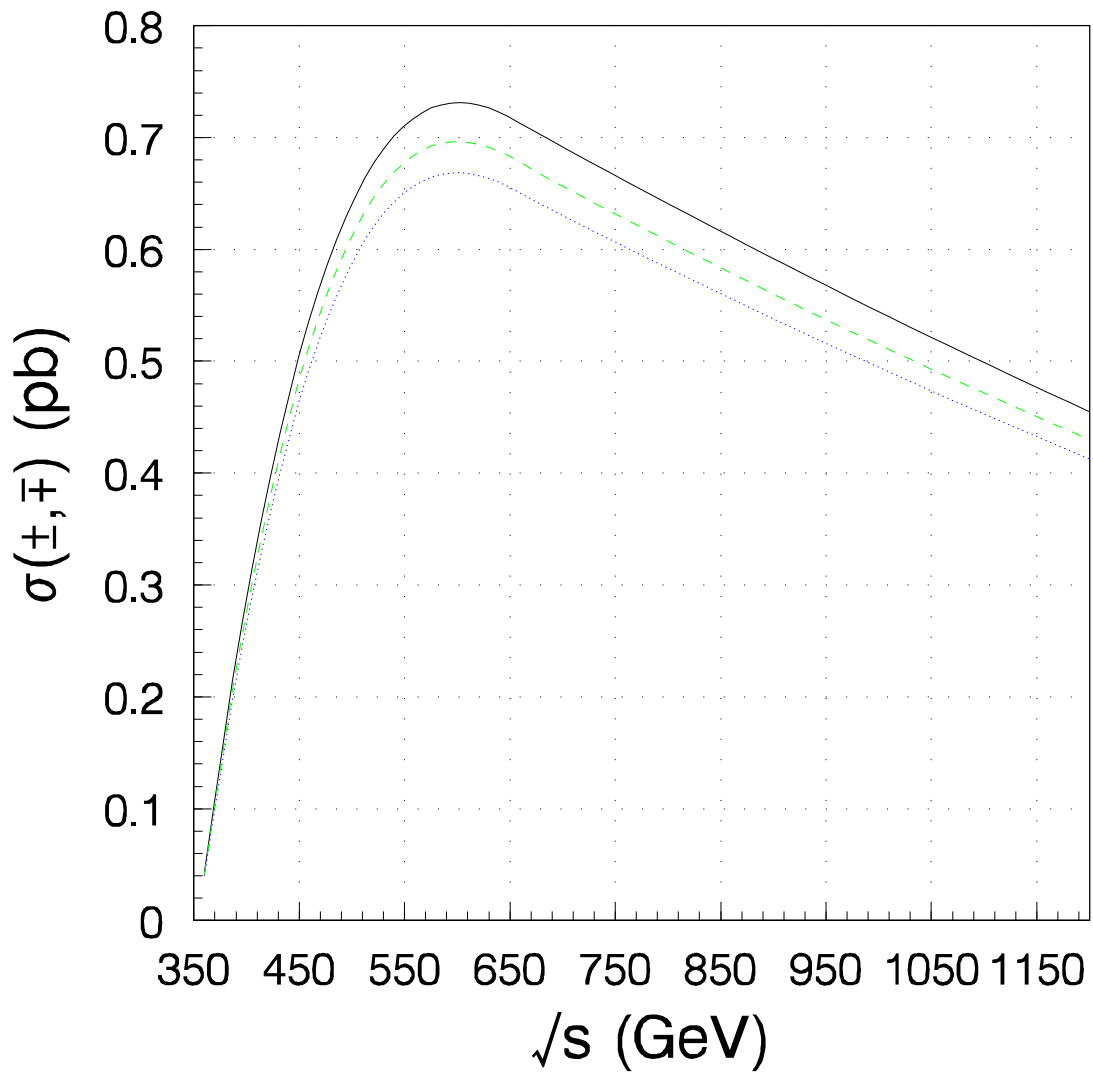
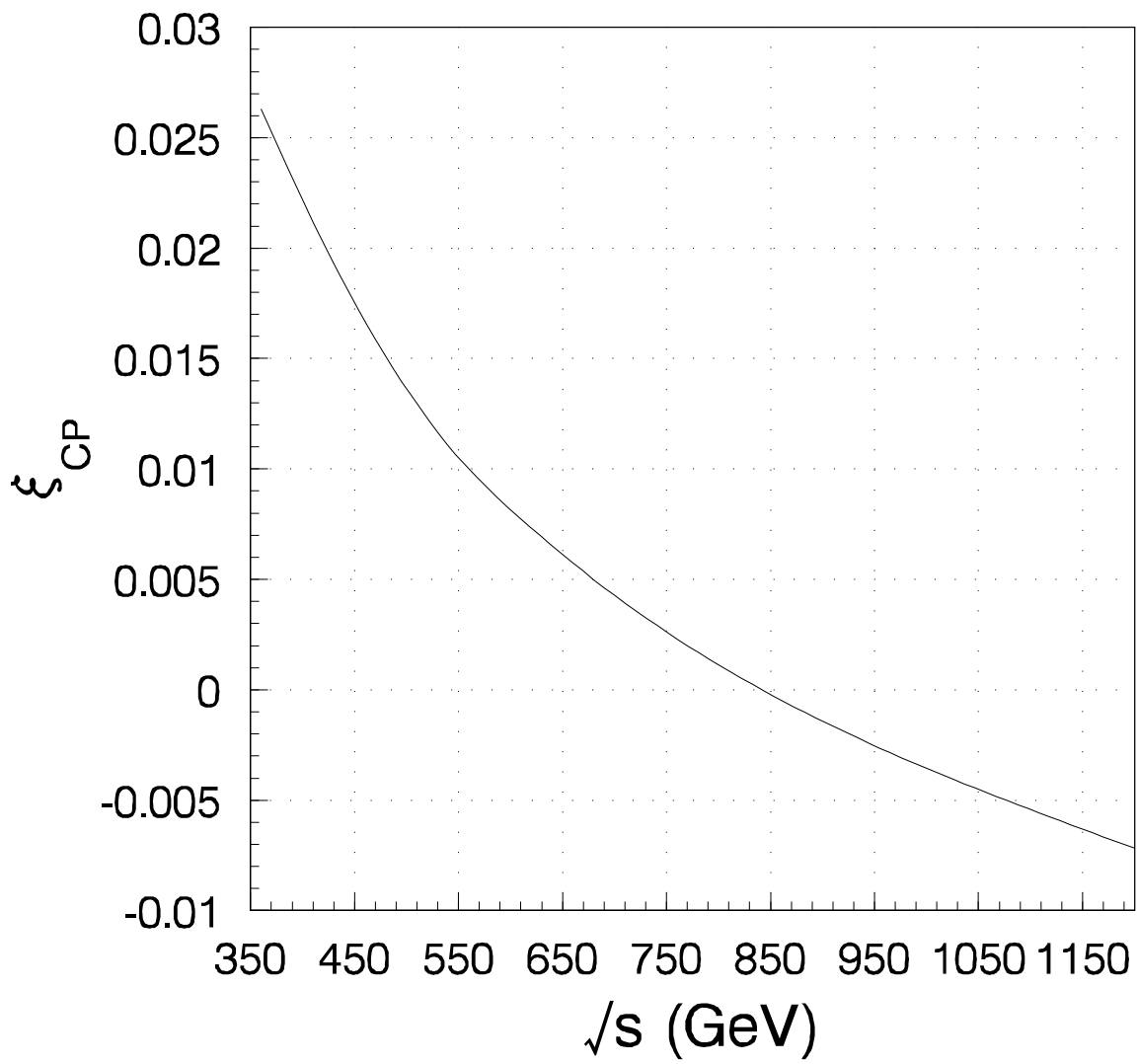
{\bf Fig.2} With $m_{\tilde{g}}=m_{\tilde{t}_1}=100~GeV$ and
$m_{\tilde{t}_2}=450~GeV$:
(a) the cross section of the $t\bar{t}$ production process $\sigma(\pm \pm)$ as
    a function of $\sqrt{\hat{s}}$,
    solid line for tree-level contribution,
    dotted line for the MSSM correction with $\phi=0^{\circ}$,
    dashed line for the MSSM correction with $\gamma(+) \gamma(+)$
    polarization and $\phi=45^{\circ}$,
    dash-dotted line for the MSSM correction with $\gamma(-) \gamma(-)$
    polarization and $\phi=45^{\circ}$;
(b) the cross section $\sigma(\pm \mp)$ as a function of $\sqrt{\hat{s}}$,
    solid line for tree level contribution,
    dotted line for the MSSM correction with $\phi=0^{\circ}$,
    dashed line for the MSSM correction with $\phi=45^{\circ}$;
(c) the CP-violating parameter $\xi_{CP}$ as a function of $\sqrt{\hat{s}}$.

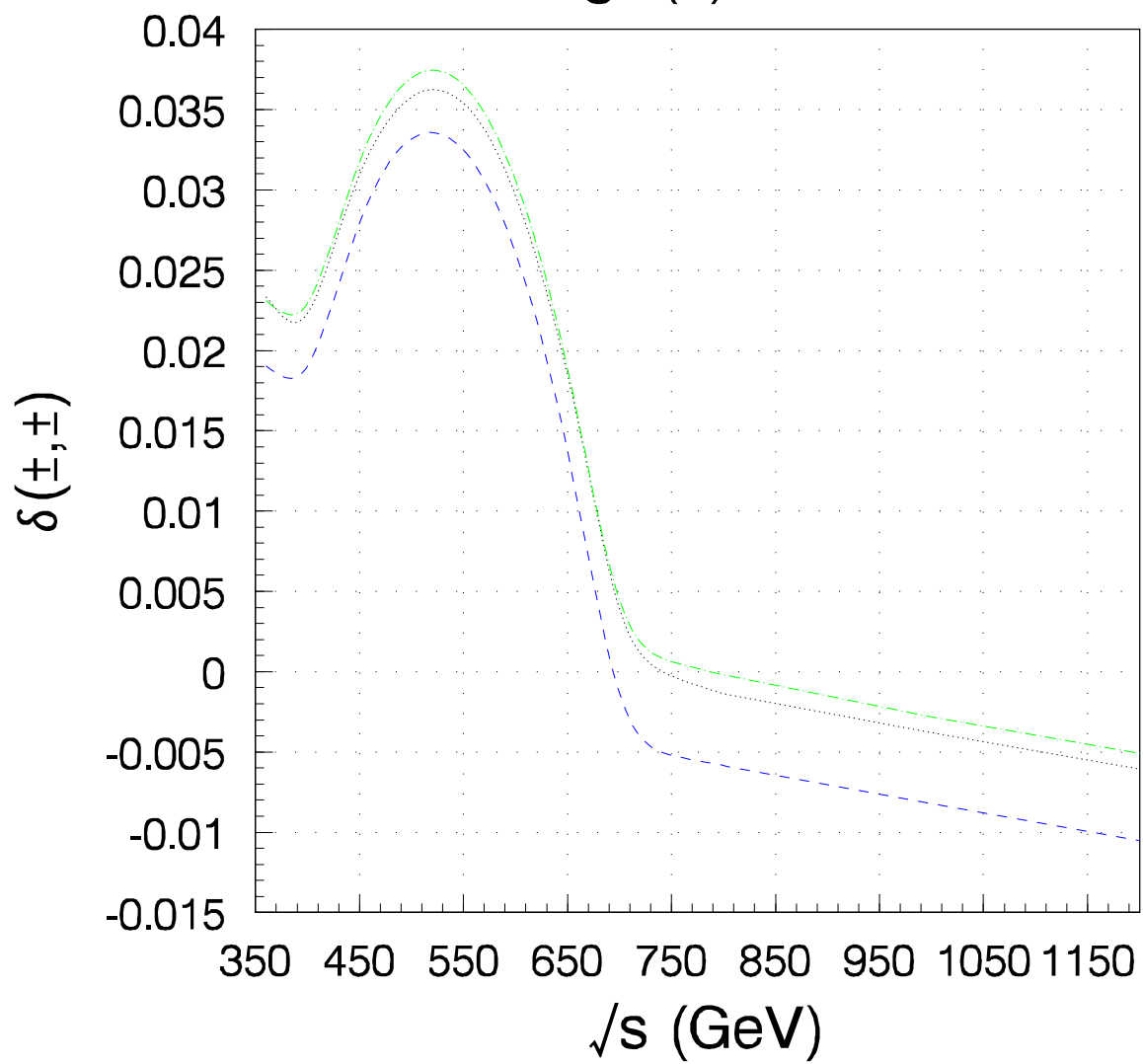
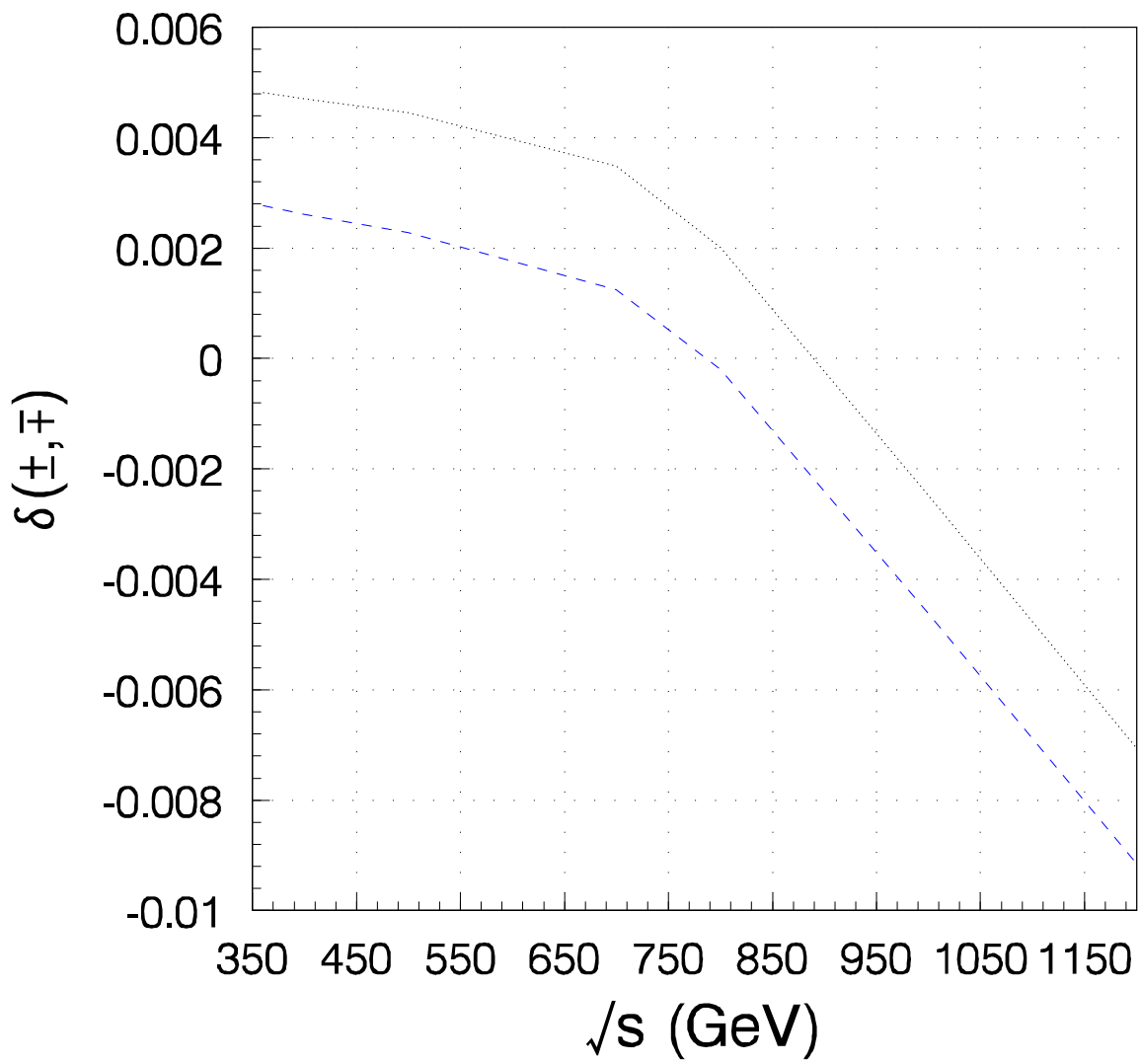
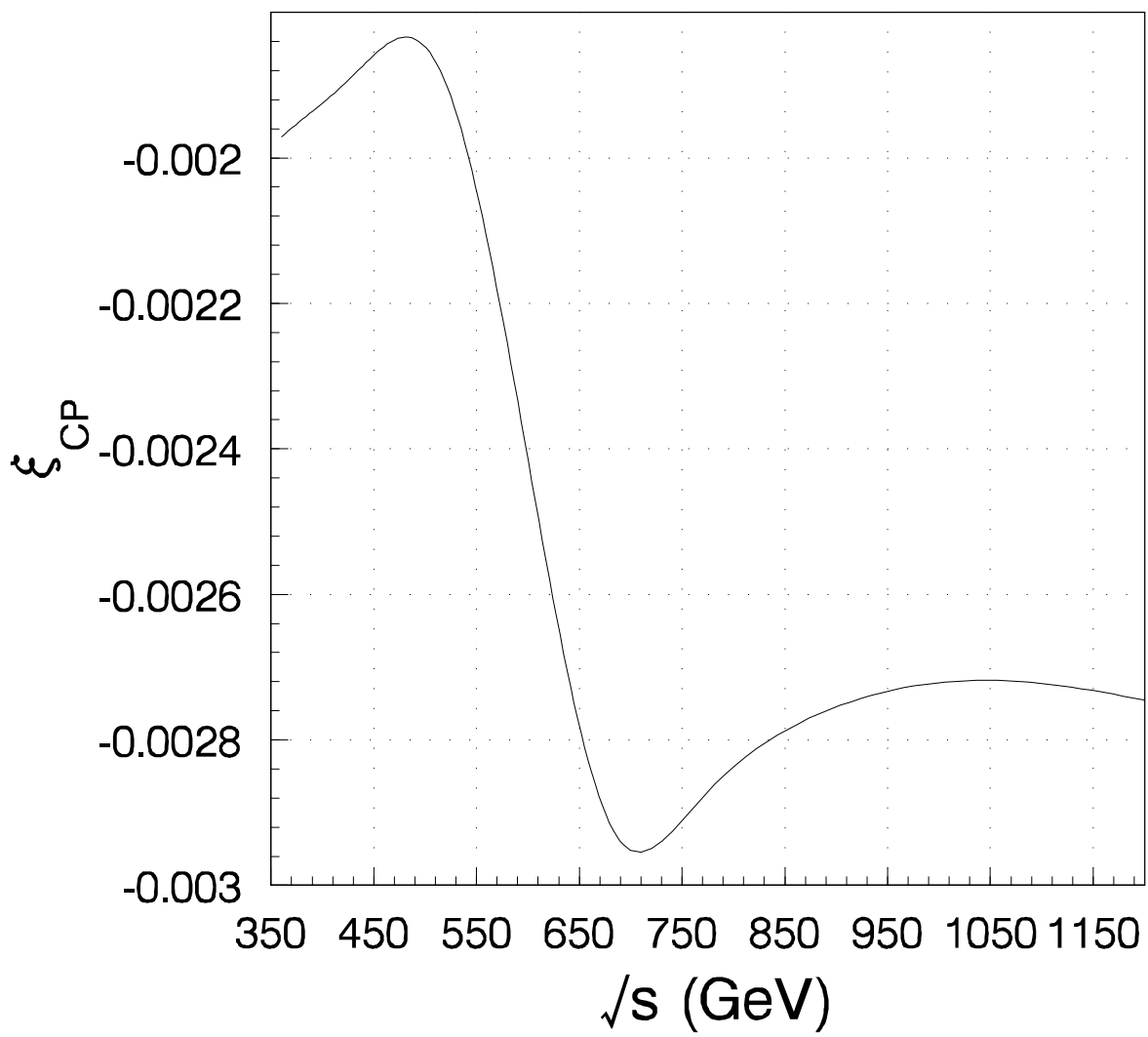
{\bf Fig.3} With $m_{\tilde{g}}=5~GeV$, $m_{\tilde{t}_1}=50~GeV$ and
$m_{\tilde{t}_2}=250~GeV$:
(a) the relative corrections to the cross section of the $t\bar{t}$ production
    process $\sigma(\pm \pm)$ as a function of $\sqrt{\hat{s}}$,
    dotted line for the MSSM correction with $\phi=0^{\circ}$,
    dashed line for the MSSM correction with $\gamma(+) \gamma(+)$
    polarization and $\phi=45^{\circ}$,
    dash-dotted line for the MSSM correction with $\gamma(-) \gamma(-)$
    polarization and $\phi=45^{\circ}$.
(b) the relative corrections to the cross section $\sigma(\pm \mp)$ as a
    function of $\sqrt{\hat{s}}$,
    dotted line for the MSSM correction with $\phi=0^{\circ}$,
    dashed line for the MSSM correction with $\phi=45^{\circ}$;
(c) the CP-violating parameter $\xi_{CP}$ as a function of $\sqrt{\hat{s}}$.

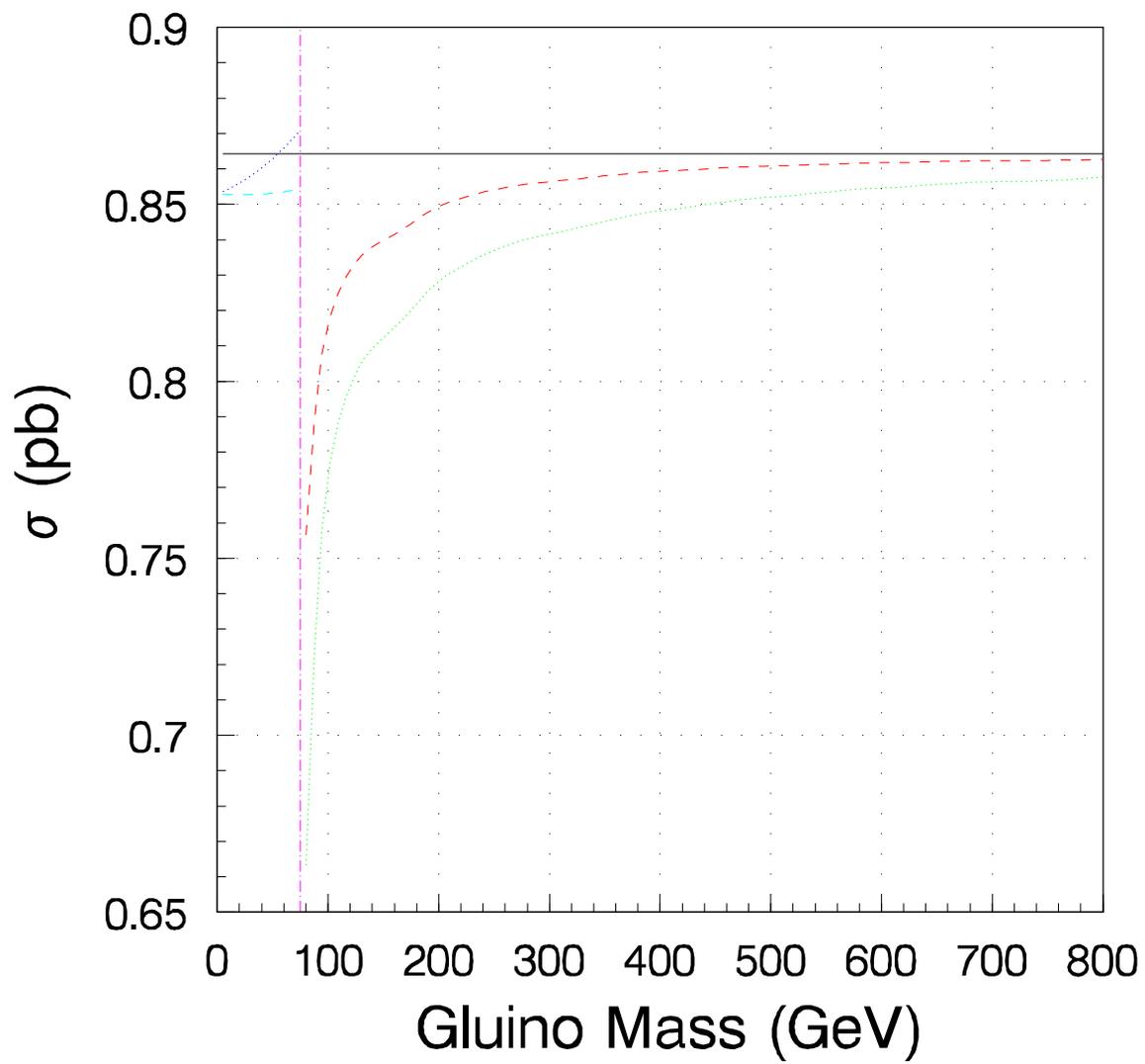
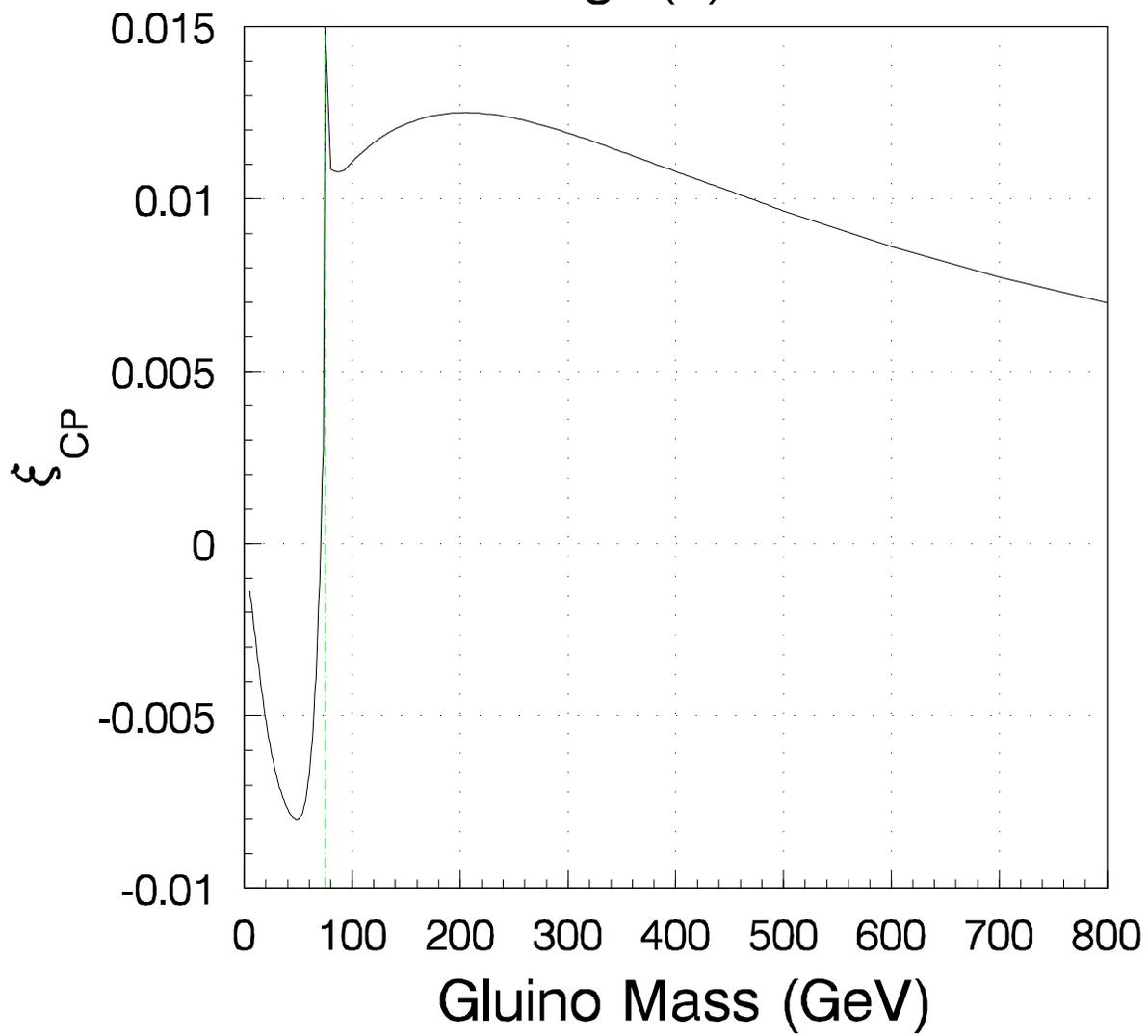
{\bf Fig.4} With $m_{\tilde{t}_1}=100~GeV$, $m_{\tilde{t}_2}=450~GeV$ and
$\sqrt{\hat{s}}=500~GeV$:
(a) the cross section $\sigma$ of the $t\bar{t}$ production process from
    unpolarized photon collision as a function of $m_{\tilde{g}}$,
    solid line for tree level contribution,
    dotted line for the MSSM correction with $\phi=0^{\circ}$,
    dashed line for the MSSM correction with $\phi=45^{\circ}$;
(b) the CP-violating parameter $\xi_{CP}$ as a function of $m_{\tilde{g}}$.

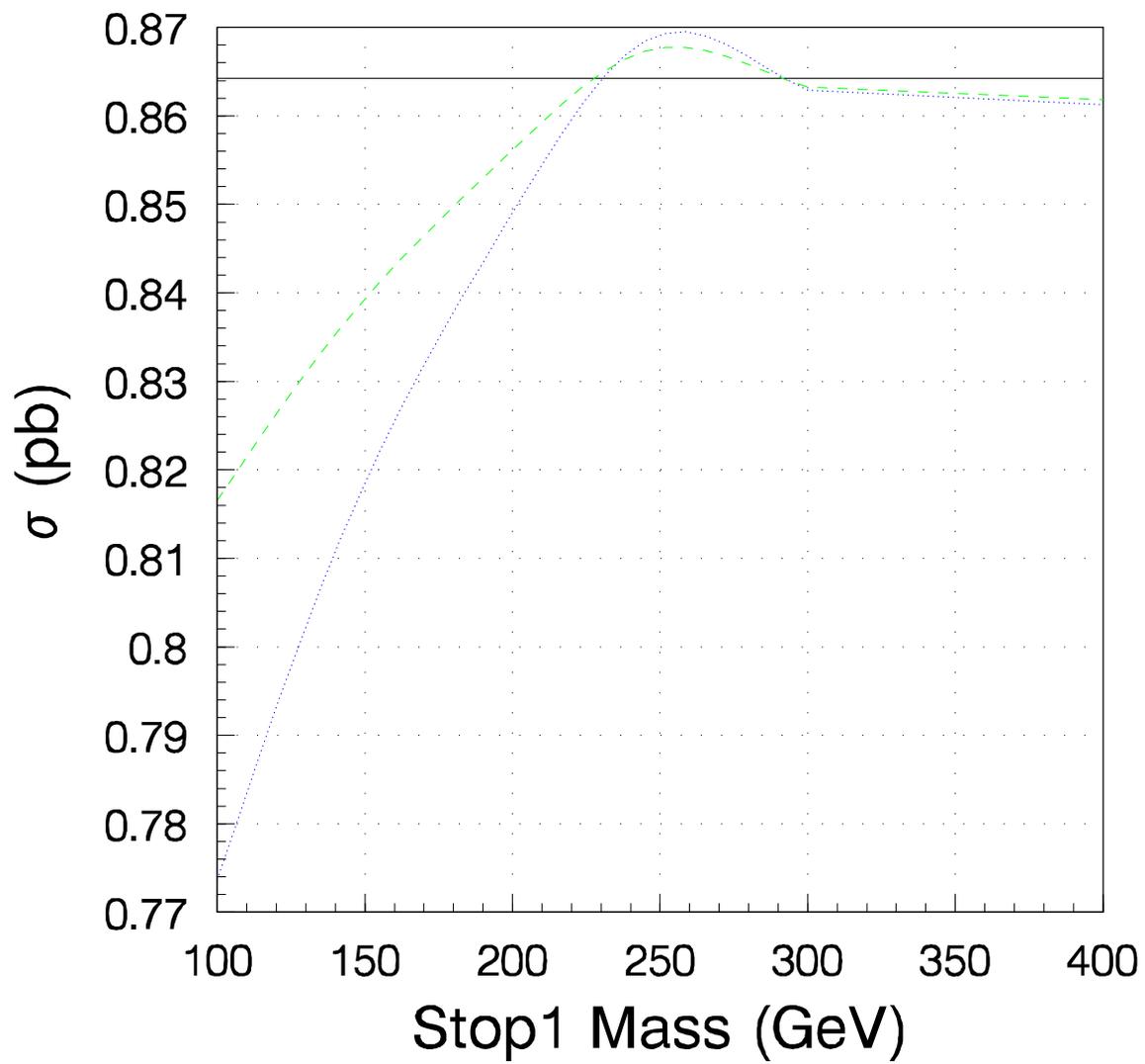
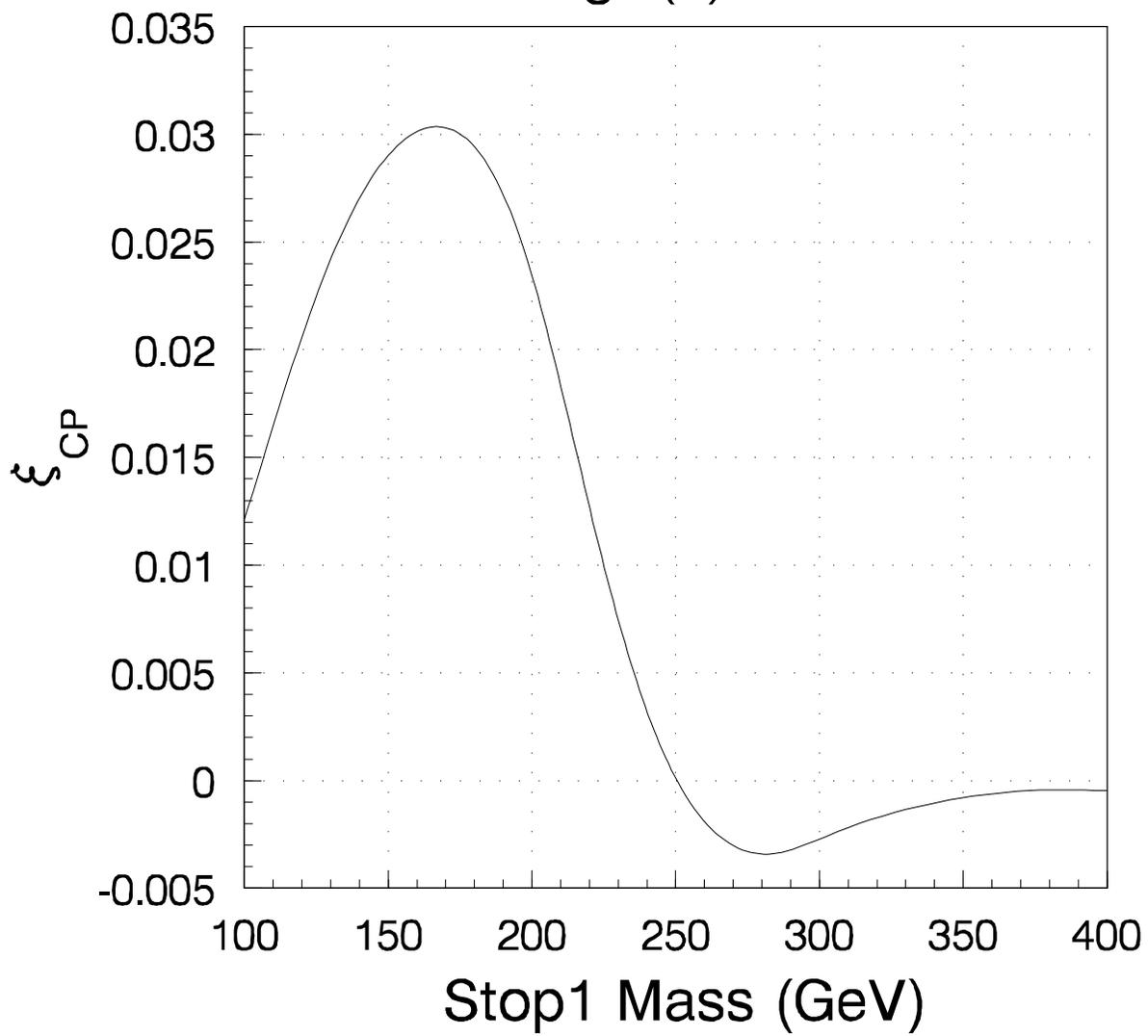
{\bf Fig.5} With $m_{\tilde{g}}=100~GeV$, $m_{\tilde{t}_2}=450~GeV$ and
$\sqrt{\hat{s}}=500~GeV$:
(a) the cross section $\sigma$ of the $t\bar{t}$ production process for
    unpolarized photon collision as a function of $m_{\tilde{t}_1}$,
    solid line for tree level contribution,
    dotted line for the MSSM correction with $\phi=0^{\circ}$,
    dashed line for the MSSM correction with $\phi=45^{\circ}$;
(b) the CP-violating parameter $\xi_{CP}$ as a function of $m_{\tilde{t}_1}$.

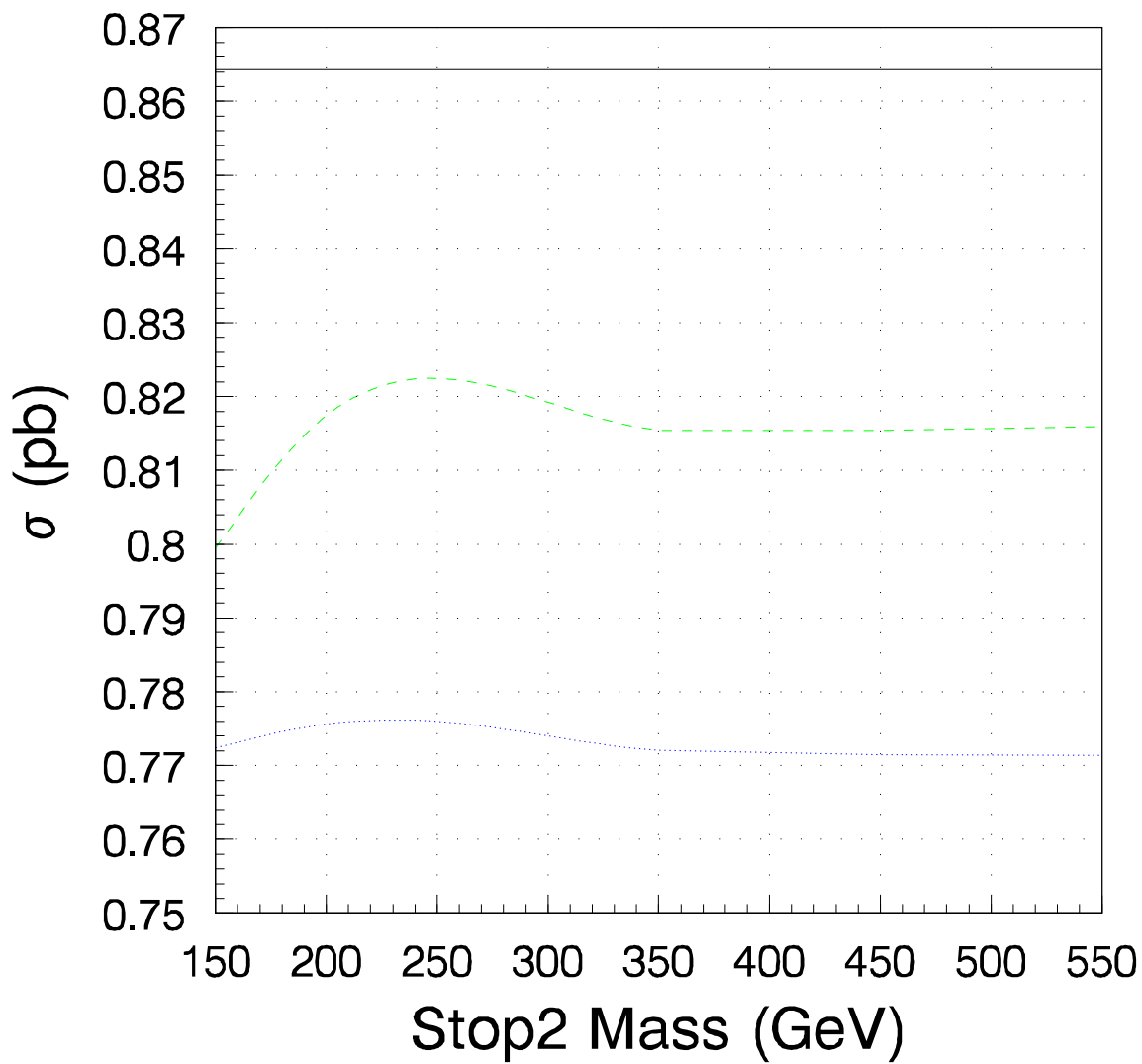
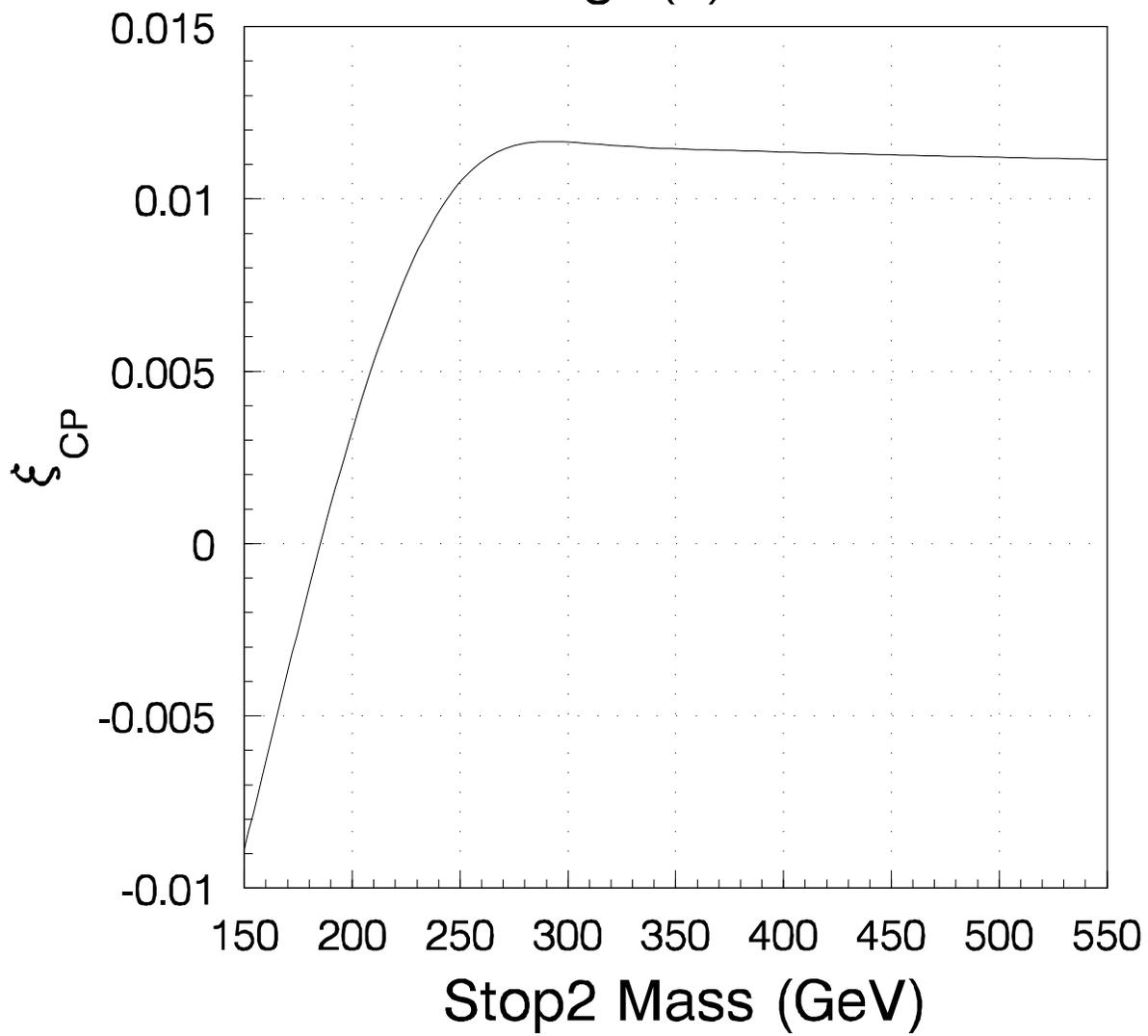
{\bf Fig.6} With $m_{\tilde{g}}=m_{\tilde{t}_1}=100~GeV$ and
$\sqrt{\hat{s}}=500~GeV$:
(a) the cross section $\sigma$ of the $t\bar{t}$ production process from
    unpolarized photon collision as a function of $m_{\tilde{t}_2}$,
    solid line for tree level contribution,
    dotted line for the MSSM correction with $\phi=0^{\circ}$,
    dashed line for the MSSM correction with $\phi=45^{\circ}$;
(b) the CP-violating parameter $\xi_{CP}$ as a function of $m_{\tilde{t}_2}$.

\end{large}
\end{document}